\documentclass[useAMS,usenatbib]{mn2e}
\usepackage{amssymb}
\usepackage{graphicx}
\usepackage[usenames]{color}
\usepackage{url}
\usepackage{enumerate}

\newcommand{\de} {{\rm d}}

\begin{document}

\title[Uncertainties and gap size in OG models]{An assessment of the pulsar outer gap model.\\
I: Assumptions, uncertainties, and implications for the gap size and the accelerating field} 

\author[D. Vigan\`o et al.]{Daniele Vigan\`o$^1$, Diego F.~Torres$^{1,2}$, Kouichi Hirotani$^3$ \& Mart\'in E.  Pessah$^4$\\ 
$^1$Institute of Space Sciences (CSIC--IEEC), Campus UAB, Faculty of Science, Torre C5-parell, E-08193 Barcelona, Spain\\
$^2$Instituci\'o Catalana de Recerca i Estudis Avan\c{c}ats (ICREA) Barcelona, Spain\\
$^3$Academia Sinica, Institute of Astronomy and Astrophysics (ASIAA), PO Box 23-141, Taipei, Taiwan\\
$^4$Niels Bohr International Academy, Niels Bohr Institute, Blegdamsvej 17, DK-2100, Copenhagen \O, Denmark}

\date{}
\maketitle

\label{firstpage}

\begin{abstract}
The popular outer gap model of magnetospheric emission from pulsars has been widely applied to explain the properties observed in $\gamma$-rays. However, its quantitative predictions rely on a number of approximations and assumptions that are usually overlooked. Here we examine them, reviewing the main ingredients entering in the model, evaluating their range of uncertainties. Usually, in the quantitative applications of the model, key parameters like the radius of curvature and the energies of the interacting photons are taken to be a fixed, single value. Instead, here we explore their realistic ranges, and the impact of these on the consistency of the model itself. We conclude that the popular evaluation of the trans-field size of the gap as a function of period and period derivative, is unreliable and affected by a huge dispersion. Last, the exploration of the possible values for the radius of curvature, the local magnetic field and other quantities deserve more attention for quantitative applications of the outer gap model, like the calculation of $\gamma$-ray spectra, which is the subject of an accompanying paper.
\end{abstract}

\section{Introduction}\label{sec:intro}

Neutron stars (NSs) are compact, spinning remnants of supernova explosions, surrounded by a region populated by plasma embedded in strong electromagnetic fields, i.e., the magnetosphere. The study of the latter has been of great interest since the discovery of the first pulsars, because the radiation processes there contribute to the detected spectrum from radio to $\gamma$-rays. For instance, the thermal X-ray radiation emitted from the surface is often overwhelmed by magnetospheric radiation (like in the Crab pulsar), or it is a minor contributor to the total energy. Among the isolated NSs, only the magnificent seven (\citealt{turolla09}) and the central compact objects (\citealt{gotthelf13}) are thought to show purely thermal emission in X-rays, with no detected contribution from the magnetosphere.
Indeed,
the large non-thermal electromagnetic fluxes we detect from pulsars are a strong indication that particles are accelerated to ultra-relativistic velocities by some mechanism, propagate in strong electromagnetic fields and lose energy in form of curvature and synchrotron high-energy photons. Particles are bound to move along magnetic field lines, therefore some region of the magnetosphere needs to host a relatively strong electric field parallel to the magnetic field lines, $E_\parallel$, that is strong enough to accelerate particles into ultra-relativistic energies. These ultra-relativistic particles efficiently emit $\gamma$-rays via synchro-curvature or inverse-Compton processes. The emitted $\gamma$-rays, in turn, should materialize as pairs via one-photon or two-photon pair-production processes, so that the charged particles may be continuously supplied in the magnetosphere.

While the basic mechanism is identified, the main concerns in modeling this emission regard the location and size of the region where particles are accelerated. In the gap models, the magnetosphere is thought to be globally force-free, with a finite electric potential within a confined, small region (referred to as the gap).
Different kinds of gap models have been developed; their details and predictions largely depend on some assumptions regarding the location, the magnetic intensity and geometry, the electrodynamics, the source of photons, and the dynamics of the pairs created. The polar cap (PC) model \citep{sturrock71,ruderman75,arons79,daugherty96} places the accelerating region close to the surface, where the radio emission is thought to be generated (\citealt{kramer97,kijak98}, see however \citealt{venter12}). The slot gap (SG, \citealt{arons83,muslimov03}) and the two-pole caustic models \citep{dyks03} extend the emission at higher altitudes, while the outer gap (OG) models \citep{cheng86a,cheng86b,romani96,hirotani99a,hirotani99b} place the gap only in the outer magnetosphere. The common properties of all gaps is that they consider rotation as the source of energy powering the gap activity. In the gaps, the electric field has a magnetic-field-aligned component, which converts part of the rotational energy into kinetic energy of particles.

In PC models, $\gamma$-rays and radio photons are generated in the same region, whereas in the SG and OG, they are not. When compared with the observed pulse profiles, outer magnetospheric models are favoured on the basis of their emission geometry \citep{romani95,romani10,johnson14,pierbattista14}, especially because radio and $\gamma$-ray pulse profiles are usually not coincident, implying different emitting regions. Furthermore, the PC models predict a super-exponential cut-off in the $\gamma$-ray spectra, due to the efficient magnetic absorption close to the surface. This is not observed \citep{1fpc,2fpc,story14}. Recently, \cite{rubtsov14} performed a blind search of $\gamma$-ray pulsars in Fermi data and analyzed the statistics of the radio-quiet and radio-loud populations. They also concluded that the outer magnetospheric models are in much better agreement with data, as compared with PC models. Finally, the recent population synthesis studies of $\gamma$-ray pulsars \citep{takata11,pierbattista12} favour the OG. 

A possible alternative to the putative existence of these gaps is represented by the striped wind models \citep{coroniti90,lyubarsky01,petri11}, in which the high-energy radiation mainly originates from the wind beyond the light cylinder. The most recent numerical studies of the global magnetospheric configuration in pulsars seem to favour such scenario, as shown by both resistive MHD simulations \citep{kalapotharakos12,kalapotharakos14} and particle-in-cell simulations \citep{chen14}. According to these works, the emission does not come from a localized region with a strong departure from force-free (a gap), but, instead, mainly from equatorial current sheets.  Part of the emission could come as well from the boundary between closed and open magnetic field lines. Note that these models are still in their infancy and the discussion about the location of the $\gamma$-ray production is still open.

In this paper, we focus on the OG model first proposed by \cite{cheng86a}, and its subsequent modifications.  It has been developed by \citet{chiang92,chiang94} and extended to the study of older pulsars, like Geminga or millisecond pulsars \citep{zhang97,cheng00,zhang02,zhang03,zhang07}, and highly magnetized pulsars \citep{takata10,tong11}.

The OG model and its extensions have become popular thanks to the quantitative estimates regarding the pulsar detectability and the gap size. Such estimates can be, in principle, analytically inferred by directly-observed pulsar timing properties, $P$ and $\dot{P}$ \citep{zhang97}. However, they are based on several hypotheses, important simplifications, and use critical model parameters.
Despite the fact that the literature on the subject is very extensive, most of these parameters and assumptions are kept fixed and unquestioned, with little exploration of the impact of these bona-fide uncertainties, leading to a lack of an exploration of the model vulnerability.

In this work, we aim to provide such systematic assessment. In \S\ref{sec:og}, we summarize the basis and the ingredients of the thin and thick OG models. We estimate the ranges of parameters in the model, relying also on Appendix~\ref{sec:magnetosphere} for issues related to the magnetospheric configuration, and on \cite{paper0} for formulae regarding the radiative losses by synchro-curvature radiation. In \S\ref{sec:efield_f} we discuss the applicability of the widely used formula of the parallel electric field as a function of spin period and surface magnetic field, both of which are supposed to constrain the trans-field size of the OG. In \S\ref{sec:discussion} we discuss the results and draw our conclusions. In an accompanying paper, we apply this study of uncertainties to the observables, evaluating, in particular, how they affect the predicted $\gamma$-ray spectrum, which is the main topic of the accompanying paper (\citealt{paper2}, Paper II).

\section{The thin and thick OG models}\label{sec:og}

\subsection{Location and geometry}\label{sec:location}

Gaps are assumed to be limited by precise boundaries between the force-free region and the gap itself, where a non-zero electric field parallel to the magnetic field line, $E_\parallel$, accelerates the particles. Ideally, the exact boundaries of the gap could be determined by considering the pair production rate, the mean free path of photons, and the local and global geometry of the magnetic field. 

The location and shape of the OG is largely dependent on the assumed configuration of the magnetic field lines. As an unperturbed state, the magnetosphere is usually treated as force-free, because the magnetic energy dominates that of the plasmas. However, to ensure enough charge supply that is needed to realize the force-free condition, there must exist a gap from which copious $\gamma$-ray photons that are capable of materializing into pairs are emitted. We can solve such a gap as a perturbation to the force-free magnetosphere. In the unperturbed state, charges co-rotate with the magnetic field lines by $\boldmath{E}_\perp \times \boldmath{B}$ drift motion, where the co-rotational electric field is defined by 
\begin{equation}
  \vec{E}=-\frac{\vec{v}_{rot}}{c}\times \vec{B} = - \frac{\vec{\Omega}\times\vec{r}}{c}\times \vec{B}~.\label{eq:efield_induced}
\end{equation}
In this unperturbed state, the real charge density, $\rho$, should coincide with the so-called Goldreich-Julian charge density,
\begin{equation}
  \rho_{gj} = -\frac{\vec{\Omega}\cdot\vec{B}}{2\pi c(1-(\Omega r \sin\theta/c)^2)}~,\label{eq:rho_gj}
\end{equation}
so that the electric field may not arise along the magnetic field lines. However, if $\rho$ deviates from $\rho_{\rm GJ}$ in some region of the magnetosphere, the force-free condition breaks down there and $E_\parallel$ appears as a perturbation. Therefore, the location and the size of the gap is essentially determined by the Poisson equation for the electro-static potential with non-vanishing $\rho - \rho_{\rm Gj}$ in some region under appropriate boundary conditions. In Appendix~\ref{sec:magnetosphere} we briefly review the magnetospheric configuration as analytically assumed (vacuum dipole or split monopole), and as obtained in numerical, global, force-free simulations. As we discuss in the accompanying paper (\citealt{paper2}, Paper II hereafter), the position of the gap, and the inner boundary in particular, strongly affects the curvature losses and, consequently, the final observable spectra in $\gamma$-rays.

\begin{itemize}

\item{\it Shape.} In its original version \citep{cheng86a}, the OG is thin in the trans-field direction, elongated along the meridional magnetic field line separating the open and closed lines, i.e., the separatrix. The meridional trans-field extension of the gap (the height, for brevity)  is assumed to be much smaller than its length (along-field extension), and it is approximated as a slab. 

\item {\it Lower boundary.} It is assumed to be the separatrix, usually calculated as the dipolar field line that is tangential to the light cylinder. The qualitative reason for this choice is geometrical: in the open field line regions, curvature photons are emitted towards the convex side of the line. As a consequence, any gap located in the convex side of some other open field line would be immediately filled by pairs and destroyed. If no high-energy photons are produced in the closed, co-rotating magnetosphere (which lies in the concave side of all open lines), the separatrix is the only region threaded by open field lines in which the curvature photons will not enter (see Figs. 3 and 4 of \cite{cheng86a}). This relies on the existence of a potential (i.e., current-free) closed field line region in the magnetosphere. If, instead, meridional currents are circulating in the co-rotating magnetosphere efficiently emitting high-energy photons, the latter could invade and destroy the gap by means of pair production.

\item {\it Upper boundary.} It is assumed to coincide with a magnetic field line. In all analytical and numerical works \citep{zhang97,takata06,hirotani06}, a fundamental related parameter is the {\it trans-field fractional size of the gap} (also referred to as the gap size/thickness), $f$, ($h_m$ of \citealt{hirotani06}), defined as:
\begin{equation}\label{eq:def_f}
 f\equiv \frac{\theta_c-\theta_u}{\theta_c}
\end{equation}
where $\theta_c$ and $\theta_u$ are the magnetic colatitudes of the footpoints on the PC surface for the lower boundary (.e., the separatrix) and the upper boundary, respectively. $f\rightarrow0$ represents a vanishingly thin gap, while $f=1$ means that the gap embraces all open field lines (from the separatrix to the rotation axis). The parameter $f$ is a fundamental regulator of the gap activity in the thick OG \citep{zhang97}, and it is assumed to be constant along the line-direction in the gap.

\item {\it Inner boundary.} The null-charge surface, where $\rho_{gj}\propto \vec{B}\cdot\vec{\Omega}=0$, Eq.~(\ref{eq:rho_gj}), is seen as a natural place to limit a gap. For an aligned rotator, the position is given by Eq.~(\ref{eq:null_sep}). Thus, the shape of the gap is mostly determined by which portion of the open field lines is crossed by the null surface. The inclination angle and the geometry of the magnetic field affect the inner boundary of the gap.

\item {\it Outer boundary.} In general, the outer boundary is very uncertain, due to the lack of knowledge about the magnetic field configuration close to the light cylinder. Analytically, the approximated Poisson equation shows that $E_\parallel$ vanishes where $d(\rho/B-\rho_{gj}/B)/dx\rightarrow 0$. Far from the surface, $\rho/B$ is constant because pairs are not created there. Thus, the outer boundary can be approximately located in the region where the gradient of $\rho_{gj}/B\sim B_z/B$ vanishes. It corresponds to the inflection point, where the field lines change from convex to concave; it is usually located within the light cylinder.

\item{\it Azimuthal boundaries.} Most of the existing gap models are 2D, with the azimuthal extension of the gap fixed to $\Delta\phi =2\pi$, or treated as a model parameter which basically regulates the flux \citep{cheng00}. Within the assumptions of the OG, the location of the gap is defined by the surface of the separatrix. For an aligned rotator, $\Delta\phi =2\pi$. When inclined dipoles are considered, the location of the null surface depends on $\phi$. As a consequence, for large inclination angles, the gap is longer in one hemisphere than in the other. \cite{wang11} consider a (simplified) three-dimensional treatment of the Poisson equation.

\end{itemize}

\subsection{The gap mechanism}\label{sec:gap_mechanism}

The basic mechanism sustaining a gap relies on the acceleration of particle by a finite $E_\parallel$.  Since the emitting particles are relativistically moving out of the gap (outwards or towards the stellar surface), a mechanism is needed to feed the necessary number of particles (the so-called {\em closure of the gap}). The production of $e^\pm$ pairs plays such a role, together with the possible extraction of particles from the surface, which depends on the poorly-known work function of electrons and the cohesive energies of ions at the NS surface. Pairs can be produced by two channels: photon-photon interaction, or photon-magnetic field interaction. The latter (magnetic channel) is effective only for strong magnetic fields, i.e., close to the surface (see, e.g., \citealt{takata10}). Otherwise, the pair production is thought to be given by the interaction between $\gamma$-ray photons produced by the accelerated particles and different kinds of ambient photons.


\begin{figure*}
\centering
\includegraphics[width=\textwidth]{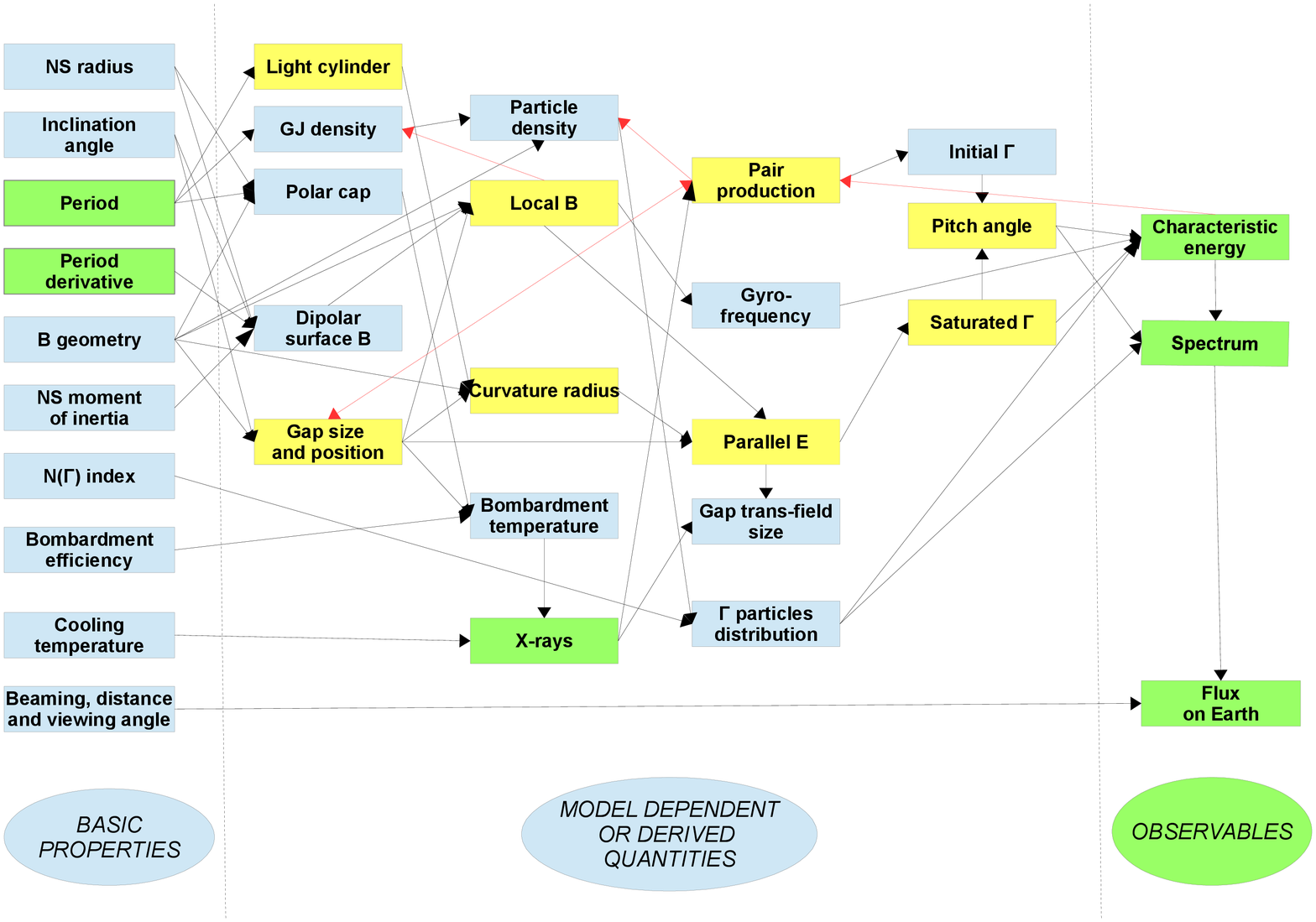}
\caption{Flow-chart of the elements and quantities defining an OG model. Yellow boxes indicate the elements for which the spectrum is more sensitive, green boxes are the observables. The order of columns is an indication of what are the most basic ingredients (leftmost part), and which other quantities are derived. The more to the right are the columns, the larger are the explicit and implicit dependencies on model parameters and assumptions. Conversely, blue arrows indicate dependencies of ingredients of the left on quantities on the right, which implies complex interdependencies between such elements.}
 \label{fig:flowchart}
\end{figure*}

We consider here a gap sustained by the X-ray flux emitted from the surface. \cite{zhang97}, and subsequent works, extend the original thin gap to the case of a thick OG, i.e., a magnetosphere which open field lines are largely depleted of particles. A big gap (i.e., large $f$) is necessary to explain the detected large $\gamma$-ray middle-age pulsars, in which the rotationally induced electric field is expected to be relatively small, due to the large periods.

In Fig.~\ref{fig:flowchart} we summarize the constituents in the $\gamma\gamma$-sustained thin/thick OG model and their mutual relations, feedbacks, and output. The physical components and the assumptions underlying them are discussed in detail in the rest of this work. 

A consistent gap model has to guarantee the pair production by the interaction between $\gamma$-ray photons and the X-ray photons. Thus, the X-ray flux becomes an important ingredient. At the same time, particle dynamics plays a central role. Particles are subject to electric acceleration and radiative losses and, as shown in \cite{hirotani99a,paper0}, they soon reach a steady state. At any given point, particles are approximately mono-energetic, i.e., they have the same Lorentz factor $\Gamma$, and emit photons. Considering this bootstrap mechanism, \cite{zhang97} propose a way to estimate the trans-field gap size, $f$, as a simple function of $P$ and $\dot{P}$, as follows.

\begin{itemize}
\item The energy of the X-ray photons is fixed at a single value for $E_X=3 kT$, which represent the frequently-interacting photons of the Wien tail of a blackbody with temperature $T$. Such surface temperature can be determined by an estimate of the intense pair bombardment onto the polar cap surface \citep{halpern93,zhang97}. They also consider the possibility of a second thermal component, extended over the entire surface, caused by the back-reflection of hot spot thermal photons by means of resonant Compton scattering. \cite{zhang02} consider $E_X$ as the peak energy of the magnetars X-ray spectra, which includes both thermal and non-thermal contributions. \cite{hirotani13} considers the temperature expected from long-lasting internal residual heat (cooling models). 
 \item For a given $E_X$, they consider a minimum energy for the $\gamma$-ray photons $E_\gamma$, according to the pair production cross-section.
 \item Assuming the curvature radiation as the only radiative process, the characteristic $\gamma$-ray photon energy, $E_\gamma$, can be associated to a Lorentz factor $\Gamma$, for a given curvature radius $r_c$.
 \item Assuming that the particles flow in the radiation-reaction steady state (i.e., radiative losses exactly compensate the electric acceleration), $\Gamma$ is associated to a unique value of $E_\parallel$, for a given rate of energy loss (for curvature radiation, it depends only on $r_c$).
 \item In the simplified treatment of the electrodynamics in the thin OG model, a recipe for the accelerating electric field is provided by electrodynamics (see \S\ref{sec:electrodynamics}), with simple dependence on the gap size $f$ and the surface magnetic field $B_\star$, $E_\parallel(B_\star,P,f)$. Applying such formula to a thick OG, $f$ becomes a simple function of $P$ and $B_\star$ only (see \ref{sec:f}). Since $f<1$ by definition, the estimate in principle predicts the $\gamma$-ray detectability of pulsars, i.e., a death-line in the $P-\dot{P}$ diagram (where the observables $\dot{P}$ and $P$ provide an estimate for $B_\star$).
 \end{itemize}


There are a number of issues related to these assumptions, namely:

\begin{itemize}
\item The estimate for the X-ray thermal flux relies on a simplified model originally proposed by \cite{halpern93}, accounting for the particles bombarding the polar caps. We discuss this issue in \S\ref{sec:xrays}.
 \item The X-ray spectral distribution with which $\gamma$-ray photons interact is broad, but it is reduced to a single $E_X$. The photon-photon cross-section depends on the angle between photons, which, in turn, depends on the local geometry (distance, direction of the magnetic field line, etc. ) and on the photon mean free path. As a consequence, the $\gamma$-ray photon have a broad dispersion around $E_\gamma$. We discuss this in \S\ref{sec:photon_energies} and Paper II.
 \item Similarly, the $\gamma$-ray energies emitted by an accelerated particle are a continuum of values. Therefore, the relation between $E_\gamma$ and $\Gamma$ is not unique.
 \item The relationship adopted between $\Gamma$ and $E_\parallel$ supposes a steady state of the accelerated particles, which could not hold if the electric field is too weak. In the steady state, radiative losses have to include the self-consistent synchro-curvature radiation, thus the Lorentz factor, in principle, depends on the gyration radius (i.e., on $B$), besides the radius of curvature (see \citealt{cheng96,paper0}). The steady state value of $\Gamma$ assumed by \cite{zhang97} takes into account only the curvature radiation.
 \item The constraints on $f$ inferred by $E_\parallel$ strongly depend on the assumed functional form of $E_\parallel$, taken as constant and uniform. As we will discuss in \S\ref{sec:efield_f}, the last estimate relies on an extrapolation to the thick OG of a simplified treatment valid for a thin OG, which implies several assumptions and limitations. For thick OG, with $f\gtrsim$ 0.1, several assumptions of the thin OG model fail, and the problem is formally more complicated. In this sense, the application of the $E_\parallel$ formula seems inconsistent with a thick OG. 
\end{itemize}

\subsection{Particle dynamics and radiative losses}\label{sec:dynamics}

The power radiated by a particle moving along around a curved magnetic field line and, at the same time, spiraling around it is \citep{cheng96,paper0}
\begin{equation}\label{eq:power_synchrocurv}
 P_{sc} = -\frac{2e^2 \Gamma^4 c}{3 r_c^2} g_r~,
\end{equation}
where $\Gamma\equiv(1-\beta^2)^{-1/2}$ is the Lorentz factor, $r_c$ is the radius of curvature of the particle trajectory, and we have introduced the synchro-curvature correction factor
\begin{equation}\label{eq:def_gr}
 g_r =  \frac{r_c^2}{r_{\rm eff}^2}\frac{[1 + 7(r_{\rm eff}Q_2)^{-2}]}{8 (Q_2r_{\rm eff})^{-1}}~,
\end{equation}
where 
\begin{eqnarray}\label{eq:q2_simple}
 Q_2^2 &&= \frac{\cos^4\alpha}{r_c^2}\left[1 + 3\xi  + \xi^2 + \frac{r_{\rm gyr}}{r_c}\right]~,\\
 r_{\rm eff} && = \frac{r_c}{\cos^2\alpha}\left(1 + \xi+ \frac{r_{\rm gyr}}{r_c}  \right)^{-1}~,\\
 r_{\rm gyr} &&= \frac{\Gamma v_\perp m c}{eB} =  \frac{\Gamma \beta\sin\alpha m c^2}{eB}~,\\
 \xi && = \frac{r_c}{r_{\rm gyr}}\frac{\sin^2\alpha}{\cos^2\alpha} \simeq 5.9\times 10^3~ \frac{r_{c,8} B_6 \sin\alpha}{\Gamma_7 \cos^2\alpha} ~.
\end{eqnarray}
$\alpha$ is the pitch angle, and hereafter $A_x=A/10^x$ in cgs units. If $\xi\ll1$, then $g_r=1$ and the losses are well described by pure curvature radiation, i.e., the radiation emitted by a particle moving exactly along the magnetic field line. Large values of $\xi\gtrsim 1$ due to, e.g., strong $B$, lead to stronger losses, due to the spiraling around the line.

\subsection{X-rays}\label{sec:xrays}

The OG model including particle bombardment \citep{halpern93} was thought to explain the two-temperature surface distribution inferred from X-ray data of the Geminga pulsar. However, additional thermal and non-thermal contributions are expected, for instance, the long-term cooling powered from the residual/Joule heat inside the NS, and the resonant Compton scattering (RCS) of thermal photons, invoked to explain the hard tail in soft and hard X-ray spectra of magnetars.

\subsubsection{Residual heat by cooling}\label{sec:cooling}

\begin{figure}
\centering
\includegraphics[width=.45\textwidth]{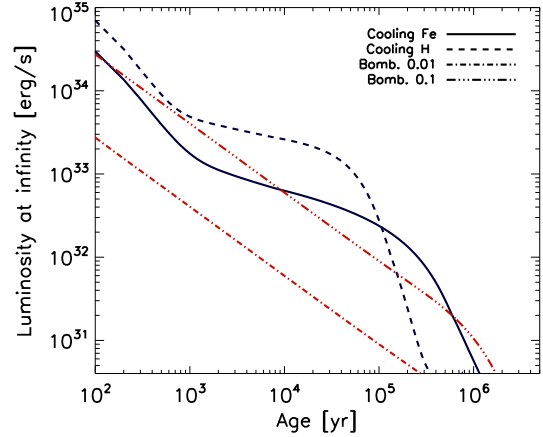}
\caption{Luminosity as predicted by cooling theory  for Iron envelope (black solid), accreted envelope (black dashed), or by particle bombardment, eq.~(\ref{eq:lx_halpern_ruderman}), with a pre-factor $\epsilon_{\rm hit}\frac{\Delta\phi}{2\pi} f(1-f/2)M_e R_6^3 k_{\rm geo}^{1/3}=0.01$ (red dot-dashed) and 0.1 (red triple dot-dash).}
\label{fig:cool_bomb}
\end{figure}
NSs are born hot ($T_0\sim 10^{10}-10^{11}$ K), then they cool down by massive neutrino emission from the core and the crust (dominating the cooling up to $\sim 10^5$ yr), and photon surface emission (dominating the cooling at later times).

In Fig.~\ref{fig:cool_bomb} we plot the evolution of the bolometric luminosity for four cooling models, taken from \cite{vigano12,vigano13}, to which we refer for all the details about the magneto-thermal evolution of isolated NSs. The models include the Ohmic dissipation and Hall effect, which drive the evolution of the magnetic field. The Joule dissipation and the anisotropic, temperature-dependent thermal and electrical conductivity are the main responsible for the interplay between the magnetic field and the temperature. Moreover, the electromagnetic torque provided by the large-scale, dipolar magnetic field affects the evolution of the spin period of the star:
\begin{equation}\label{eq:ppdot_spindown}
  P \dot{P} = K_{\rm sd} B_p^2~,
\end{equation} 
with
\begin{equation}\label{eq:k_spindown}
  K_{\rm sd}=h_\iota\frac{2\pi^2}{3}\frac{R_\star^6}{Ic^3}=  2.44\times 10^{-40}~ h_\iota \frac{R_6^6}{I_{45}} \mbox{~s G}^{-2}~,
\end{equation}
where $I = I_{45} 10^{45}$ g cm$^2$ is the moment of inertia of the star, and $h_\iota \sim O(1)$ is the inclination angle-dependent factor uncertainty in the spin-down formula (see e.g. \citealt{spitkovsky06} and \S2.2 of \citealt{viganothesis} for a detailed discussion). In literature, the fiducial values $I_{45}=1$, $R_6=1$ and $h_\iota=1$ (vacuum orthogonal rotator) are commonly used.

The black solid and dashed lines represent the model with Iron envelope, or accreted envelope, respectively, with an initial dipolar magnetic field $B_0=10^{13}$ G (see model A of \cite{vigano13} for further details). We note that the theoretical uncertainty on the cooling luminosity can span about one-two orders of magnitude, depending on the NS mass, superfluidity model, equation of state. Strong magnetic fields, $B\gtrsim 10^{14}$ G, lead to an enhanced luminosity by orders of magnitude, due to the dissipation of crustal currents.

The theoretical temperature suffers from the further uncertainty of which emission model is chosen (blackbody, atmosphere of different chemical composition). The predicted temperature at $\sim 10^3-10^4$ yr can vary up to a factor $\sim 3$.  Moreover, the surface temperature is thought to be inhomogeneous due to the anisotropic conductivity induced by a strong magnetic field \citep{perna13}. Emitting radii smaller than the surface generally agree with what can be inferred from the spectral fits. As a rule of thumb, we expect that young NSs $t\lesssim 10^5$ yrs have $T\gtrsim 10^6$ K, and that the most magnetized NSs are hotter, due to the extra heat deposited in the crust by Joule dissipation. However, from a purely theoretical point of view, given the mentioned uncertainties, it is not possible to associate a temperature to a given set of values $(P,B_\star)$. In the absence of detected X-ray thermal flux, the surface temperature can be treated as a free parameter within the range expected from cooling theory.

Relying on 1D cooling models and the classical thick gap formula for the trans-field gap size (with the inclusion of a screening factor, see \S~\ref{sec:f}), \cite{hirotani13} discusses how the residual heat affects the evolution the $\gamma$-ray luminosity: during the first $\sim 10^4$ it is almost constant, then it decreases slower than the rotational energy, so that the efficiency rises.

Last, note that internal heat can power X-ray thermal emission during the first millions of years at most. Therefore, the thermal emission of millisecond pulsars, typically of the order of $0.5-0.7$ MK, can only be explained by particle bombardment, with a radius of a few km which is roughly consistent with the expected polar cap radius, Eq.~(\ref{eq:pc_radius}). As we see below, the value of the temperature can constrain some physical parameter related to the particle bombardment.

\subsubsection{Particle bombardment}\label{sec:bombardment}

The pairs created in the OG are accelerated in opposite directions due to $E_\parallel$ in the gap. Then, the particles propagating inward will exit the gap and radiate part of their energy in their way to the NS surface (where $E_\parallel \simeq0$) and will deposit the remaining energy on the polar cap, heating it up. The latter subsequently radiates thermal X-rays \citep{halpern93}.
The amount of energy that the pairs will deposit upon hitting the stellar surface can be estimated by equating the rate of change in their kinetic energy to the radiative losses by synchro-curvature radiation, Eq.~(\ref{eq:power_synchrocurv}):
\begin{equation}
\label{eq:motion_polar_cap}
\dot \Gamma m c^2 = -\frac{2c}{3} \left(\frac{e}{r_c}\right)^2 g_r\Gamma^4~.
\end{equation}
The equation of motion along the line ($d_l= - ct$),
\begin{equation}\label{eq:ekin_int}
 - 3 \int_{\Gamma_{\rm in}}^{\Gamma_\star}\frac{d\Gamma}{\Gamma^4} = - \frac{2e^2}{mc^2} \int_{d_{\rm in}}^{R_\star} \frac{g_r}{r_c^2}(d_l) ~\de d_l ~,
\end{equation}
can be integrated from the inner limit of the OG (for a given $d_{\rm in}$ and $\Gamma_{in}$), to the stellar radius, $R_\star$:

\begin{equation}\label{eq:gamma_surface}
\Gamma_\star = \left[ \frac{1}{\Gamma^3_{\rm in}} + \frac{2e^2 k_{\rm sc} }{mc^2 R_{lc}}\right]^{-1/3}~,
\end{equation}
where we have defined the dimensionless geometrical factor
\begin{equation}
 k_{\rm sc} = R_{lc} \int_{R_\star}^{d_{\rm in}} \frac{g_r}{r_c^2}(d_l)~\de d_l  ~.
\end{equation}
In order to evaluate Eq.~(\ref{eq:gamma_surface}), we have to consider the geometry of the line, which fixes the dependences on $d_l$, the position of $d_{\rm in}$ and the value of $\Gamma_{\rm in}$. In particular, the synchro-curvature factor $g_r/r_c^2$ plays an important role in slowing down the particles. We refer to the Appendix Eqs.~\ref{sec:distance_line} and \ref{sec:rc} for details about the distance along the line and the radius of curvature; there we show that close to the surface, $d_l\sim r$.

Analytical estimates can be done as follows. The value of $d_{\rm in}$ can be taken as the location of the null surface for the separatrix in a vacuum dipole, Eq.~(\ref{eq:null_general}). It depends on the spin period, and on the inclination and azimuthal angles. Luckily, $\Gamma_\star$ is almost insensitive to the value of $d_{in}$, because most of the particle energy is radiated close to the surface, where the radius of curvature is smaller.

Analytical formulae for $r_c(r,\theta)$ are possible for the separatrix in a vacuum non-rotating dipole (see Appendix~\ref{sec:magnetosphere}. Close to the surface, it can be approximated by $r_c=k_c(d_l R_{lc})^{1/2}$, with $k_c=4/3$ (see Eq.~(\ref{eq:rcsep})). With the additional, important hypothesis that $g_r=1$ (purely curvature radiation), and $\Gamma_{\rm in} \gg \Gamma_\star$, then $k_{sc}=\ln(d_{\rm in}/R_\star)/k_c^2$, and Eq.~(\ref{eq:gamma_surface}) simplifies to
\begin{eqnarray}\label{eq:gamma_surface_simple}
\Gamma_\star & = & \left[\frac{4 \pi e^2}{mc^3 k_c^2 P} \ln\left(\frac{d_{\rm in}}{R_\star}\right)\right]^{-1/3} \nonumber \\
& = & 1.3\times 10^7 ~k_{\rm geo}^{1/3} P^{1/3}~,
\end{eqnarray}
where we have expressed $R_{lc}$ in terms of the spin period, $R_{lc}=cP/2\pi$ (hereafter $P$ is implicitly given in units of seconds), and
\begin{equation}
 k_{\rm geo} = \frac{r_c}{(d_lR_{lc})^{1/2}}\frac{1}{\ln(d_{\rm in}/50 R_\star)}  ~.
\end{equation}
\cite{halpern93} and \cite{zhang97} take $k_c=1$, and an arbitrary, $P$-independent value $d_{\rm in}=50~R_\star$ (i.e., $k_{\rm geo} = 1$). 

In general, the kinetic energy of the particles impacting the surface is
\begin{equation}
 E_{\rm kin \star}= \Gamma_\star  m_e c^2~.
\end{equation}
With the approximation (\ref{eq:gamma_surface_simple}), we obtain
\begin{equation}\label{eq:energy_HR93}
E_{\rm kin\star}\simeq 10.4~k_{\rm geo}^{1/3} P^{1/3}~{\rm erg }~.
\end{equation}
On the other hand, if $\Gamma_{\rm in}$ is too small (so that synchro-curvature losses are negligible), then $\Gamma_\star \simeq \Gamma_{\rm in}$, and
\begin{equation}\label{eq:ekin_gin}
  E_{\rm kin\star}= E_{\rm kin}(d_{\rm in}) = 8.2~\Gamma_{\rm in, 7}~{\rm erg }~,
\end{equation}
where all the possible dependences (e.g., with $P$) are hidden in the value of $\Gamma_{\rm in}$.

\begin{figure*}
\centering
\includegraphics[width=.45\textwidth]{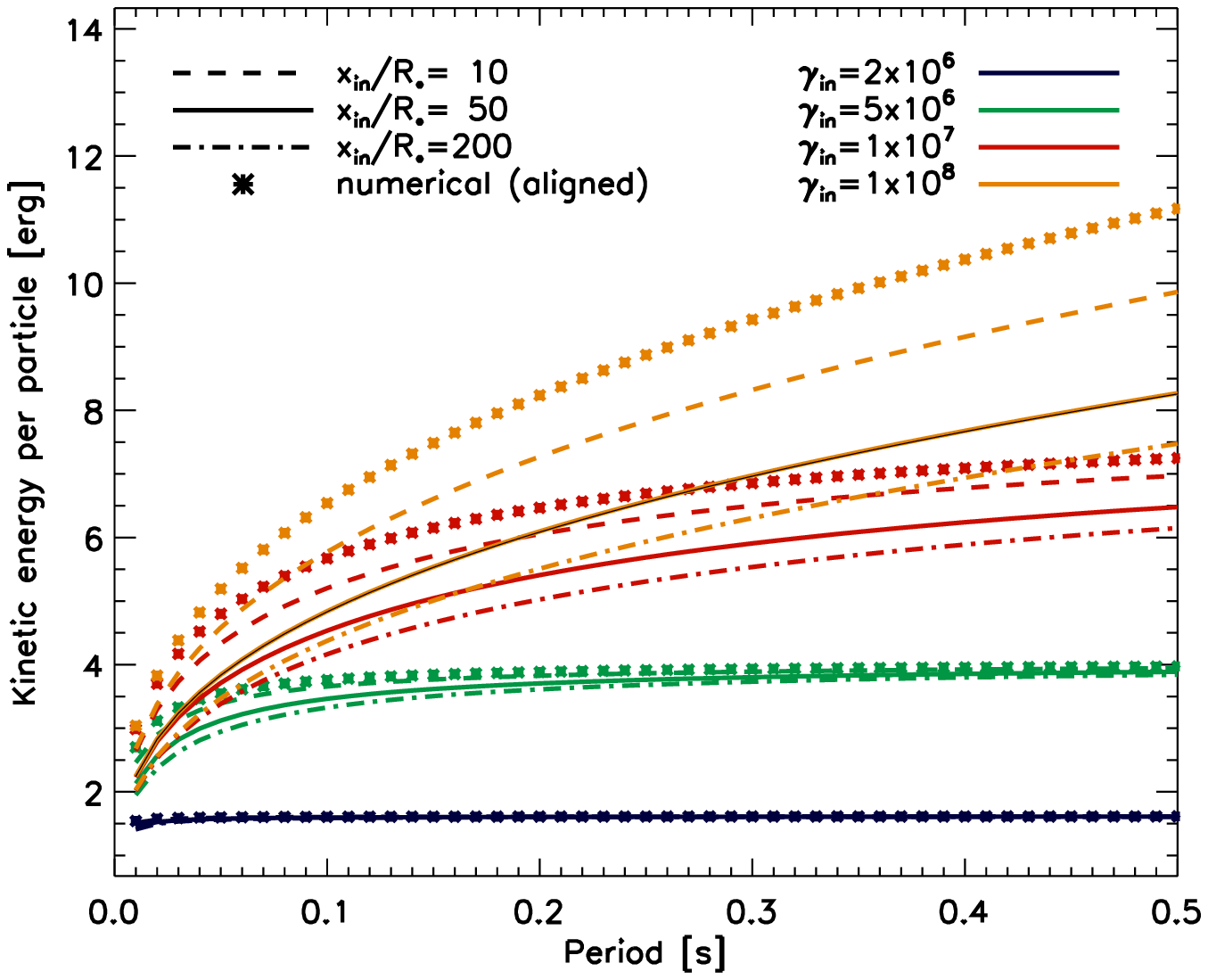}
\includegraphics[width=.45\textwidth]{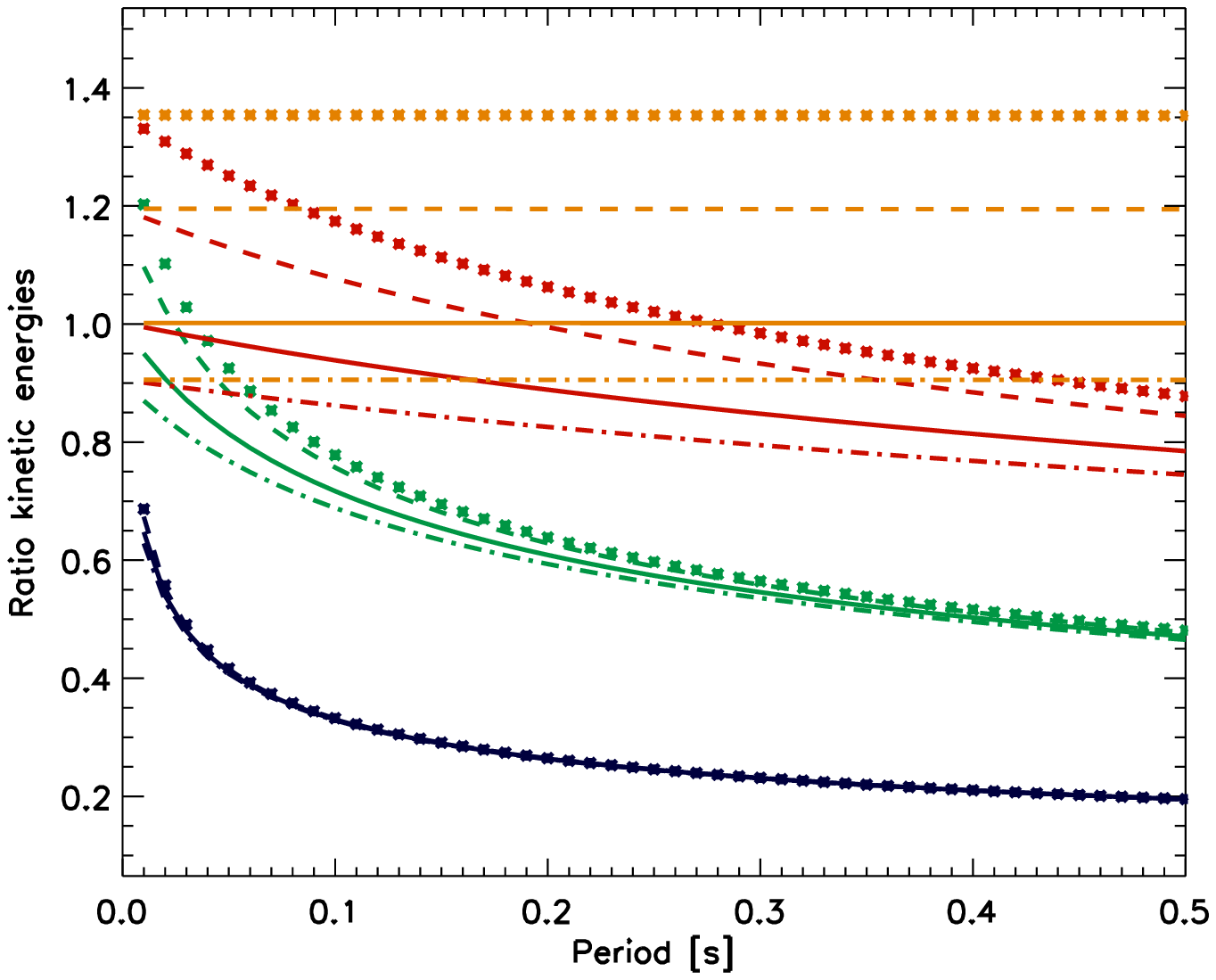}
\caption{Kinetic energy per particle hitting the polar cap, according to Eq.~(\ref{eq:gamma_surface}), for different choices of $\Gamma_{\rm in}$ (colors). Asterisks indicate the solutions obtained by numerical integration of Eq.~(\ref{eq:gamma_surface}), while lines stand for different fixed values of $x_{in}/R_\star$. Left: values of kinetic energy. Right: ratio between the calculated energies and the formula (\ref{eq:gamma_surface_simple}) (thin black line in the left panel).}
 \label{fig:ekin}
\end{figure*}

In Fig.~\ref{fig:ekin}, we confirm these estimates by showing the results of the numerical integration of Eq.~(\ref{eq:ekin_int}) for an aligned vacuum dipole, as a function of $P$, for different values of $\Gamma_{\rm in}$ and fixed $R_\star=10$ km. We employ the numerical function $r_c(d_l)$ for the non-rotating dipole separatrix, Fig.~\ref{fig:distance_line}, and assume purely curvature radiation, $g_r(d_l)=1$. We adopt $d_{\rm in}/R_{lc}=2/3$, roughly corresponding to the intersection between the null surface and the separatrix for an aligned rotator (Eq.~\ref{eq:null_sep}).

We also over-plot the analytical values of $E_{\rm kin}(R_\star)$ as a function of period, by considering Eq.~(\ref{eq:gamma_surface}), with $k_c=1$ (black thin line). It coincides with the case $d_{\rm in}=50~R_\star$ and $\Gamma_{\rm in}=10^8$, since the latter is large enough. Also for lower $\Gamma_{\rm in}$ and very fast spinning pulsars, $P\lesssim 0.1$~s, the approximation holds, because the curvature radiation is efficient. The numerical and analytical calculations are compatible, with minor differences due to the choices of $d_{\rm in}$. The $P^{1/3}$ dependence reflects the fact that larger the period, larger the light cylinder, thus larger the distance along which particles travel and lose energy.

However, the approximation~(\ref{eq:gamma_surface_simple}) fails if $\Gamma_{\rm in}$ is small, and/or the period is long: in both cases, the curvature losses are not efficient enough to slow down to $\Gamma_\star \ll \Gamma_{\rm in}$, and the kinetic energy approaches the value (\ref{eq:ekin_gin}). As a consequence, the dependence with period becomes very weak. On the other hand, the choice of different values for $d_{\rm in}$ has a minor influence on the deposited energy.

The standard reference employed for OG models is Eq.~(\ref{eq:energy_HR93}), with $k_{\rm geo}=1$ \citep{halpern93,zhang97}. We stress that such formula is derived under several assumptions: 1) The movement considered is along the separatrix of a vacuum dipolar magnetic field configuration. 2) The radius of curvature is taken to be $r_c = \sqrt{d_l/R_{lc}}$, which, corrected by a factor $k_c=4/3$, holds in the regions close to the surface for a dipole, with intense curvature emission. 3) The value $d_{\rm in}/R_\star$, arbitrarily fixed to 50, this weakly affects $E_{\rm kin}(R_\star)$.

In addition, there is an effect that tends to make Eq.~(\ref{eq:energy_HR93}) an overestimation. The radiative losses are assumed to be due to curvature only, i.e., $g_r=1$. The inclusion of the synchro-curvature formulae would lead to a smaller value for $E_{\rm kin \star}$.

The fundamental conclusion about $E_{\rm kin \star}$ is that the large uncertainties (up to one order of magnitude) arise from the assumptions that the radiative losses are purely due to the curvature process, $\Gamma_{in}\gg \Gamma_\star$ and that the radius of curvature close to the surface is $R_{lc}\sim d_l^{1/2}$. With this in mind, the usual approximation is likely overestimating by a factor of a few the amount of energy deposited by the particles.

\subsubsection{Temperature of the hot spot}\label{sec:bombardment2}

From the kinetic energy deposited onto the surface, one can estimate the polar cap temperature. Depending on the sign of $E_\parallel$, either positrons or electrons that are produced as pairs in the OG, are accelerated star-wards by $E_\parallel$ and eventually strike the polar-cap (PC) surface. Let us evaluate the number flux of such in-falling particle species. The footpoints of the magnetic field lines along which they fall, will distribute on a portion of a ring-like region on the PC surface, which extends between $r_p(1-f)$ and $r_p$ in the $\theta$ direction, and $\Delta\phi$ in the azimuthal direction. As a consequence, the surface footprints of the OG subtend an area
\begin{eqnarray}\label{eq:a_og}
 A_{\rm og} &=& \frac{\Delta\phi}{2\pi}\pi r_{\rm pc}^2 [1 - (1-f)^2] \nonumber \\
                   &=& \Delta\phi f \left(1-\frac{f}{2}\right)\frac{2\pi R_\star^3}{Pc} ~,
\end{eqnarray}
where the polar cap radius, $r_{\rm pc}$, has been defined in Eq.~(\ref{eq:pc_radius}) as a function of the period, and $f$ stands for the trans-field fractional size of the gap (\ref{eq:def_f}).\cite{zhang97} use $A_{\rm og}=2f\pi r_{\rm pc}^2$, which is valid under the thin gap approximation, $f\ll 1$, and assumes that the azimuthal extension of the gap is $\Delta \phi = 2\pi$.

The particle density in and below the gap is expected to deviate from $n_{gj}=\rho_{gj}/e$, where $\rho_{gj}$ is given by Eq.~(\ref{eq:rho_gj}). Thus, we define the particle density above the surface as $n_e(R_\star)=M_en_{gj}(R_\star)$, where $M_e$ accounts for such deviations from $n_{gj}$ evaluated above the surface. Note that \cite{zhang97} use $M_e=1/2$.

Since particles flow with velocity $v=c$, the flux of particles hitting the surface is estimated as:
\begin{eqnarray}\label{eq:ne_flux}
\dot N_e & =& A_{\rm og} M_e n_{gj}(R_\star) c \nonumber \\
& \simeq &  \Delta\phi f \left(1-\frac{f}{2}\right)r_{\rm pc}^2M_e~\frac{\Omega B_\star}{2\pi e}  \nonumber \\
 & \simeq &   \Delta\phi f \left(1-\frac{f}{2}\right)M_e~2\pi \frac{R_\star^3 B_\star}{c e P^2}  \\
 & \sim & 2.7 \times 10^{30}  \left(\frac{\Delta\phi}{2\pi} f \left(1-\frac{f}{2}\right) M_e R_6^3~\frac{B_{12}}{P^2}\right) \; \textrm{s}^{-1} ~, \nonumber
\end{eqnarray}
where $R_6=R_\star/10^6$ cm. With $M_e=1/2$ and $\Delta\phi (1-f/2)=2\pi$ we recover the values by \cite{zhang97}.

The particles deposit a total energy $\dot E_{\rm hit} = E_{\rm kin\star}\dot N_e$ onto the stellar surface, where $E_{\rm kin}(R_\star)$ has been estimated above. Using Eq.~(\ref{eq:energy_HR93}), we obtain

\begin{equation}\label{eq:en_hitting}
   \dot{E}_{\rm hit} \simeq 
    2.9 \times 10^{31}\frac{\Delta\phi}{2\pi} f(1-f/2)M_e R_6^3 k_{\rm geo}^{1/3}~\frac{B_{12}}{P^{5/3}} ~{\rm erg~s}^{-1}~.
\end{equation}
If the kinetic energy is converted into thermal X-ray luminosity with efficiency $\epsilon_{\rm hit}<1$, then
\begin{eqnarray}\label{eq:lx_halpern_ruderman}
L_X \simeq \epsilon_{\rm hit} \dot{E}_{\rm hit}~.
\end{eqnarray}
Assuming a homogeneous, isotropic blackbody emission from this region, the simple relation
$L_X =  A_{\rm og}\sigma T_{\rm h}^4$ holds, where $\sigma=5.67\times 10^{-5}$ erg cm$^{-2}$ s$^{-1}$ $K^{-4}$ is the Stefan's constant. Therefore
\begin{equation}\label{eq:t_h_pre}
T_h = \left[ \frac{\epsilon_{\rm hit} E_{\rm kin\star}M_e n_{gj} c }{\sigma}\right]^{1/4} = \left[ \frac{\epsilon_{\rm hit} E_{\rm kin\star}M_e B_\star }{\sigma eP}\right]^{1/4}~.
\end{equation}
This bombardment-induced temperature, $T_h$, is not explicitly dependent on $f$, instead it is only given by the assumed geometry, particle dynamics and density. Under the assumptions made in Eq.~(\ref{eq:energy_HR93}), we obtain\footnote{To recover Eq.~(7) of \cite{zhang97}, we set $M_e=1/2$ and $k_{\rm geo}=\epsilon_{\rm hit}=1$. Note also that, in the same equation, the quoted dependence $R_6^{3/4}$ is incorrect: there is no such dependence, since the temperature is independent of $A_{\rm og}$.}

\begin{eqnarray}\label{eq:t_h}
T_h &=&  4.4\times 10^6  (\epsilon_{\rm hit} M_e)^{1/4} k_{\rm geo}^{1/12} ~ B_{12}^{1/4}P^{-1/6} ~{\rm K } \\
 &= & 0.38 (\epsilon_{\rm hit} M_e)^{1/4} k_{\rm geo}^{1/12} ~ B_{12}^{1/4}P^{-1/6} ~{\rm keV} \,.
\end{eqnarray}
On the other hand, if we consider the lower values for $E_{\rm kin \star}$, Eq.~(\ref{eq:ekin_gin}), we obtain:

\begin{equation}\label{eq:en_hitting2}
   \dot{E}_{\rm hit} \simeq 2.2 \times 10^{31}\frac{\Delta\phi}{2\pi} f(1-f/2)M_e R_6^3 k_{\rm geo}^{1/3}~\frac{B_{12} \Gamma_7}{P^2} ~{\rm erg~ s}^{-1}~,
\end{equation}
and the blackbody temperature is
\begin{eqnarray}\label{eq:t_h_pre2}
T_h = && 4.2\times 10^6  (\epsilon_{\rm hit} M_e)^{1/4} k_{\rm geo}^{1/12} ~ B_{12}^{1/4}P^{-1/4} ~{\rm K } \\
k_bT_h = && 0.36 (\epsilon_{\rm hit} M_e)^{1/4} k_{\rm geo}^{1/12} ~ B_{12}^{1/4}P^{-1/4} ~ {\rm keV} \,.
 \end{eqnarray}
Note that the temperature slightly depends on the large uncertainties of $E_{\rm kin \star}$, discussed above. Also the dependencies on period, magnetic field and geometry are weak.

The red lines of Fig.~\ref{fig:cool_bomb} represent the expected luminosity due to the particle bombardment, for the same cooling model used to calculate the cooling curve (black lines). The magneto-thermal models allow us to follow the evolution of the dipolar component of the surface $B$, which regulates the electromagnetic torque, and, consequently, the evolution of the spin period. Having these quantities, the bombardment luminosity is calculated by means of Eq.~(\ref{eq:lx_halpern_ruderman}), with the value of the pre-factor $\epsilon_{\rm hit}\frac{\Delta\phi}{2\pi} f(1-f/2)M_e R_6^3 k_{\rm geo}^{1/3}=0.01$ (dot-dashed) and 0.1 (triple dot-dashed). Note that this pre-factor is not expected to be constant through the evolution, since the gap size will vary, becoming wider for older ages. Given the dispersion of the pre-factor, and the intrinsic assumptions behind its formula (see previous section), we conclude that, similarly to the theoretical cooling curves (black lines), the luminosity given by the bombardment can be estimated within a range of a few orders of magnitude. Thus, it is not easy to say which is the dominant mechanism. Qualitatively, one expects that, at late times, the gap is larger and the pre-factor is higher, thus making the bombardment an important contribution. The large uncertainties prevent us from making firmer conclusions.

\subsubsection{Resonant Compton scattering (RCS)}\label{sec:rcs}

\cite{halpern93} proposed RCS to account for the two different temperatures seen in the X-ray spectrum of Geminga. They considered the scattering between the X-rays from the heated polar cap (see \S\ref{sec:bombardment}) and the plasma particles close (within a few stellar radii) to the polar caps. The optical depth of the process is supposed to be much larger than unity, and all the X-rays from the heated polar cap are reflected back to the star illuminating homogeneously the whole surface. The temperature of the soft X-rays satisfied $\epsilon_{\rm r}\epsilon_{\rm hit} L_X = \sigma A_{\rm r} T_s^4$, where $\epsilon_{\rm r}$ represents the fraction of outgoing photon energy which is effectively reflected back homogeneously over an area $A_{\rm r}$. Therefore the temperature of the area emitting soft X-rays is

\begin{equation}
   T_s = T_h \left(\epsilon_{\rm hit}\epsilon_{\rm r}\frac{A_{\rm og}}{A_{\rm r}}\right)^{1/4}~,
\end{equation}
where we considered the efficiency of the conversion from bombardment kinetic energy to thermal radiation, the free parameter $\epsilon_{\rm hit}$. With $A_{\rm og}$ given by Eq.~(\ref{eq:a_og}), we have
\begin{eqnarray}
\label{eq:t_s}
T_s & \simeq & 4.5 \times 10^5 {\cal K}_s ~ B_{12}^{1/4} P^{-5/12} ~{\rm K} \\
k_bT_s & \simeq & 0.039 {\cal K}_s ~ B_{12}^{1/4} P^{-5/12} ~ {\rm keV} ~,
\end{eqnarray}
where the pre-factor ${\cal K}_s$ includes all the geometrical and efficiency factors discussed until now:
\begin{equation}
 {\cal K}_s \equiv  \epsilon_{\rm hit}^{1/2} \left[\frac{\Delta\phi}{2\pi} f\left(1-\frac{f}{2}\right)R_6 \epsilon_{\rm r} M_e\frac{4\pi R_\star^2}{A_{\rm r}} \right]^{1/4} k_{\rm geo}^{1/12} \label{eq:kappa_s}
\end{equation}
Note the large uncertainties in the value of $T_s$ hidden in the pre-factor ${\cal K}_s$.\footnote{To recover Eq.~(9) of \cite{zhang97}, set $M_e=1/2$, $A_r=4\pi R_\star^2$, and $k_{\rm geo}=\epsilon_{\rm hit}=\epsilon_r=1$. They neglect the factor $(1-f/2)^{1/4}$.} There are a number of fundamental complications concerning this picture, which we list below.

\begin{itemize}
 \item {\it Doppler shift of the resonant frequency.} Particles moving at velocity $\beta$ scatter off photons shifted by a factor $\Gamma(1-\cos\theta_{\gamma B} \beta)$, 
 \begin{equation}\label{eq:redshift_frequency}
\omega_D(B,\beta)=\frac{1}{\Gamma(1-\beta\cos\theta_{\gamma e})}\frac{eB}{m_ec}~,
\end{equation}
 \item {\it Density and velocity of the scattering particles.} In order to evaluate the optical depth of RCS scattering, \cite{zhang97} use the particle density resulting from the the pairs flowing from the gap to the polar cap. They magnetically generate a cascade of pairs, which are the scatterers of the X-ray photons. However, most of these particles are likely ultra-relativistic, thus the Doppler shift is extremely important, and IR/optical photons would be scattered instead of X-rays. In the original article by \cite{halpern93} and in \cite{cheng99}, the X-ray photons scatter with the co-rotating plasma in the closed region, where particles may be only mildly relativistic. In any case, the Doppler effect has to be considered.
 \item {\it Reflection efficiency.} Reflected photons will very unlikely spread homogeneously across the whole surface, as assumed in this model. Depending on the height where the scattering takes place, the solid angle subtended by the star could be: a) too small and most photons could mostly miss the star ($\epsilon_{\rm r} \ll 1$), b) too large and reflected photons would illuminate only a smaller fraction of its surface ($A_{\rm r} \ll 4\pi R_\star^2$). Even in the most optimistic case, one would expect as a maximum value $\sim$ half of the surface. Furthermore, the temperature is not expected to be homogeneous, because the thermal flux is transported mainly radially (see, e.g., \citealt{kaminker14} for internal heat sources). Additionally, the typical magnetic fields of pulsars are strong enough to suppress the trans-field transfer of heat, enhancing the maintenance of inhomogeneities in $T_s$.
 \end{itemize}

We conclude that the RCS process on hot polar cap thermal photons is unlikely to work as described in the thick OG literature, and that the estimates of $T_s$ (e.g. \cite{zhang97}) suffer very large uncertainties. In general, the claimed unique relations between a bombardment-related temperature ($T_h$ or $T_s$) to a given set of $P,B_\star$ values are affected by large dispersions. At the same time, the long-term cooling temperature is often neglected, while in reality it can exceed the bombardment temperature and dominate the photon-photon interaction. 

Note also that RCS is thought to be very effective in converting soft thermal photons into high-energy photons in magnetars, where the magnetosphere is twisted and dense \citep{rea08,beloborodov13}. Since the process is effective close to the surface (up to $\sim 10 R_\star$, where the local $B$ and the particle density are large enough), these up-scattered X-rays will enter into the OG (if present), which lies at higher altitudes. As a consequence, the typical X-ray energy is larger than the for a purely surface, thermal emission.

\subsection{Energies of interacting photons}\label{sec:photon_energies}

Once the distribution of the X-ray flux is specified, we can evaluate the pair-production rate at each point of the magnetosphere, taking account of the collisions between these X-rays and the gamma-rays emitted from the OG. The cross section of the $\gamma$-$\gamma$ pair production, for two photons having energies $E_1$, $E_2$ is \citep{gould67}:

\begin{equation}\label{eq:pp_cross_section}
 \sigma^{\gamma\gamma}= \frac{3\sigma_T}{16} (1-\mu^2) \left[2\mu(\mu^2-2) + (3-\mu^4)\ln\left(\frac{1+\mu}{1-\mu}\right)\right]~,
\end{equation}
where we have used the Thomson cross section, 
\begin{equation}\label{eq:thomson}
  \sigma_T = \frac{8\pi}{3}\left(\frac{e^2}{m_ec^2}\right)^2 = 6.65\times10^{-25} \mbox{cm}^2~,
\end{equation}
and the ``center of mass'' parameter
\begin{equation}\label{eq:beta_pair_production}
\mu=\sqrt{1-\frac{2(m_e c^2)^2}{(1-\cos \psi) E_1 E_2}} ~,
\end{equation}
and where $\psi$ is the angle between the propagation directions for the two photons. Inverting this relation, we obtain
\begin{equation}\label{eq:e1e2}
 E_1E_2 = \frac{2}{(1-\cos\psi)}\frac{(m_ec^2)^2}{(1-\mu^2)}~.
\end{equation}
Note that head-on collisions ($\cos\psi=-1$) are strongly favored, while strictly tail-on collisions ($\cos\psi=1$) will not produce pairs, independently on the energies of the photons. The maximum value of the cross section is obtained for $\mu_{\rm max}=0.701$, which implies $(1-\mu^2_{\rm max})\simeq 0.5$, thus
\begin{equation}\label{eq:e1e2max}
 E_1E_2(\mu^{\rm max}) \simeq \frac{4}{(1-\cos\psi)}(m_ec^2)^2~.
\end{equation}
The cross section approaches zero for $\mu\rightarrow 1$, i.e., very large energies, and for $\mu=0$, i.e., the very minimum energies of the photons required for the process to be possible:
\begin{equation}\label{eq:e1e2min}
 E_1E_2(\mu=0) =\frac{2(m_ec^2)^2}{(1-\cos\psi)}~.
\end{equation}
In our scenario, we can parametrize the most likely interacting $E_\gamma$, by defining a factor $k_\gamma$ in Eq.~(\ref{eq:e1e2}):
 \begin{equation}\label{eq:egamma}
   E_\gamma = \frac{2}{(1-\cos\psi)}\frac{0.26 {\rm~ GeV}}{(1-\mu^2) E_X [{\rm keV}]}\equiv\frac{k_\gamma}{kT [{\rm keV}]} {\rm ~GeV}~.
 \end{equation}
In order to estimate the range of the X-ray energies, consider that, for any blackbody with temperature $T$, $\sim 90\%$ of the photons are emitted with energies in the range $0.15-5.3~kT$, and $\sim 50\%$ between $0.75-2.8~kT$ (the range we consider below). The peak of the photon distribution is for $E_X=1.6~kT$, but the Wein tail is more likely to interact with $\gamma$-rays, and this may justify the usual choice $E_X= 3~kT$.

The breadth of values of $\mu$, which also contains the dependence on $\psi$, can be considered by employing the cross-section in two limiting cases: $\mu=0$, and $\mu^{\rm max}=0.701$. Larger values of $\mu$ are less favoured, since they require more energetic $E_\gamma$ and, at the same time, the cross section decreases.
By considering these ranges for the variables, and the limit $\cos\psi=-1$, we can obtain the corresponding energy range for $E_\gamma$. Low values are obtained considering $\mu=0$ and energetic X-ray photons, $E_X=2.8~kT$. Larger values for $E_\gamma$ are obtained for $\mu=0.71$ and $E_X=0.75~kT$. Substituting these pairs of values in Eq.~(\ref{eq:egamma}), the range of $k_\gamma$ is $\sim [0.1,3.5]$. \cite{zhang97} employ $\mu=0$ (i.e., vanishing cross-section) and $E_X=3~kT$, corresponding to $k_\gamma=0.87$.

Note that these single-value estimates only give an idea of the expected $E_\gamma$, while proper numerical calculations, including the optical depth, are needed to study the details (see Paper II for the impact of these changes in the predicted spectra).

\subsection{Electric field needed to sustain the gap}\label{sec:electric_field}

For a given Lorentz factor, the synchro-curvature characteristic energy is given by \citep{cheng96}:
\begin{equation}\label{eq:characteristic_energy}
 E_c(\Gamma,r_c,r_{\rm gyr},\alpha) = \frac{3}{2}\hbar cQ_2\Gamma^3~.
\end{equation}
By equating $E_c$ to the characteristic energies given by Eq.~(\ref{eq:egamma}), we can estimate the needed $E_\parallel$, assuming the radiation-reaction steady state, i.e. $eE_\parallel + P_{sc} =0$, where $P_{sc}$ is the synchro-curvature power, Eq.~(\ref{eq:power_synchrocurv}):
\begin{equation}\label{eq:epar_general}
  E_\parallel  = \left(\frac{2}{3}\right)^{7/3}e\frac{g_r}{r_c^2} \left(\frac{E_\gamma}{Q_2\hbar c}\right)^{4/3}~,
\end{equation}
and $g_r$ and $Q_2$ are the factors defined in eqs.~(\ref{eq:def_gr}) and (\ref{eq:q2_simple}). Expressing the light cylinder as a function of $P$, Eq.~(\ref{eq:def_light_cyl}) and $r_c/R_{lc}$, and substituting $E_\gamma$ with Eq.~(\ref{eq:egamma}), we have
\begin{eqnarray}
  E_\parallel & =& \left(\frac{2}{3}\right)^{7/3}\frac{e}{c^2}\left(\frac{E_\gamma}{\hbar}\right)^{4/3}\tilde{g}_r\left(\frac{2\pi R_{lc}}{P r_c}\right)^{2/3} \nonumber\\
  &\simeq & 3.6\times 10^6 \tilde{g}_r \left(\frac{R_{lc}}{r_c P}\right)^{2/3}  \left[ \frac{k_\gamma}{kT [{\rm keV}]}  \right]^{4/3} \frac{{\rm V}}{{\rm m}}~, \label{eq:epar_t}
\end{eqnarray}
If we employ the temperatures derived above, Eqs.~(\ref{eq:t_h}) and (\ref{eq:t_s}), then:
%
%
%
\begin{eqnarray}
  E_\parallel (T_h) & \simeq & \frac{1.3\times 10^7}{P^{4/9} B_{12}^{1/3}}\left( \frac{k_\gamma}{{\cal K}_h}  \right)^{4/3} \tilde{g}_r \left(\frac{R_{lc}}{r_c}\right)^{2/3}~ \frac{{\rm V}}{{\rm m}} ~, \label{eq:epar_th} \\
  E_\parallel (T_s) & \simeq & \frac{2.8\times 10^8}{P^{1/9} B_{12}^{1/3} f^{1/3}} \left( \frac{k_\gamma}{{\cal K}_s}  \right)^{4/3} \tilde{g}_r \left(\frac{R_{lc}}{r_c}\right)^{2/3} ~ \frac{{\rm V}}{{\rm m}}~,\label{eq:epar_ts} 
\end{eqnarray}
with

\begin{eqnarray}
 {\cal K}_h & \equiv & (\epsilon_{\rm hit} M_e)^{1/4} k_{\rm geo}^{1/12}~, \label{eq:kappa_h} \\
 {\cal K}_s & \equiv & \epsilon_{\rm hit}^{1/2} \left[\frac{\Delta\phi}{2\pi} \left(1-\frac{f}{2}\right)R_6 \epsilon_{\rm r} M_e\frac{4\pi R_\star^2}{A_{\rm r}} \right]^{1/4} k_{\rm geo}^{1/12}~, \label{eq:kappa_s2}\\
 \tilde{g}_r & = & \frac{g_r}{(Q_2r_c)^{4/3}}~,
\end{eqnarray}
where $\tilde{g}_r=1$ in the limit of purely curvature radiation. This estimate for the electric field basically relies on a given set of local values for $r_c,B,\alpha,E_X$.

Note that this derivation relies on a specific choice of $E_c=E_\gamma$, which implies that the shape of the synchro-curvature spectrum is basically arbitrary fixed. In Paper II we face this issue.

\section{Electric field and gap size}\label{sec:efield_f}

\subsection{Assumptions behind the $E_\parallel$ prescription}\label{sec:electrodynamics}

In the gap, the force-free condition is locally violated, because the local electric charge density, $\rho_e$, differs from the $\rho_{gj}$, Eq.~(\ref{eq:rho_gj}) i.e., the value for a perfectly force-free, charge-separated plasma. We can find the non-corotational electric potential by writing the Poisson equation as

\begin{equation}\label{eq:poisson}
 \nabla^2 V^{\rm nc} = -4\pi (\rho_e - \rho_{gj})~.
\end{equation}
Boundary conditions have to be specified in order to solve the Poisson equation and describe the electrodynamics.
\cite{cheng86a} proposes a solution for a {\it thin gap}, with the following assumptions:

\begin{itemize}
\item the trans-field thickness of the gap, $a$, is very small, $a\ll R_{lc}$, implying $f\ll 1$;
\item the geometry of the gap reduces to a planar 2D slab, with coordinates along $\vec{B}$-lines, $x \ge 0$ (where $x=0$ identifies the inner boundary, but no outer boundary is fixed), and across them (trans-field direction) $z \in [0,a]$, where $a$ is the trans-field width of the gap; 3D effects (finite inclination angle, azimuthal dependences) are not considered;
\item if the inner boundary of the gap is given by the null surface, it only crosses a small fraction of the open magnetic field lines; as a consequence, $f$ cannot be close to one, by geometrical construction. The maximum value (i.e., the fraction of open lines) depends on the inclination angle: for example, we obtain $f<0.6$ for the magnetic inclination of 60 degrees, provided that the magnetic field configuration
is described by the vacuum, rotating dipole solution \citep{cheng00}.
\item the gap is assumed to be completely void of particles, $\rho_e=0$ in Eq.~(\ref{eq:poisson}); this is a good approximation only if pair production is not too efficient, so that $\rho_e \ll \rho_{gj}$ (inefficient screening);
\item the GJ density, Eq.~(\ref{eq:rho_gj}), is evaluated in the limit $r\ll R_{lc}$, for which $\rho_{gj}=- \vec{\Omega}\cdot\vec{B}/2\pi c$;
\item the electric field in the gap slightly deviates from the rotationally induced force-free one, Eq.~(\ref{eq:efield_induced}); this implies the assumption $E_\parallel \ll E_\perp$;
\end{itemize}
With these assumptions and approximations, Eq.~(\ref{eq:poisson}) simplifies to:
\begin{equation}\label{eq:poisson_gap}
 \nabla^2 V^{\rm nc} = \frac{d^2V^{\rm nc}}{d^2x} + \frac{d^2V^{\rm nc}}{d^2z} = - \frac{2}{c}\vec{\Omega}\cdot\vec{B}
\end{equation}
Additional geometrical assumptions and approximations, taken in order to find a solution, are:

\begin{itemize}
\setcounter{enumi}{6}
\item the trans-field width $a$ is constant along the field line direction;
\item the strength of the field $B$ is taken as constant in the whole gap;
\item the null surface is simplified to lie along the $z$-axis of the slab (i.e., perpendicular to $\vec{B}$), while, by definition, it lies where $\vec{\Omega}\cdot\vec{B}=0$;
\item the local angle between $\vec{B}$ and $\vec{\Omega}$ is approximated such that $\cos\theta_{B\Omega} \sim x/r_c$, where $x$ is the distance along the field line from the null surface, which is valid only if $x\ll r_c$.
\end{itemize}
Then, the Poisson equation~(\ref{eq:poisson}) reduces to an ordinary differential equation, which is solved with the following boundary conditions:

\begin{itemize}
\item $V^{\rm nc}=0$ on the upper and lower boundaries $z=0$, $z=a$;
\item $E_x^{\rm nc}=0$ at the inner boundary.\footnote{In Eq.~(3.5) of \citealt{cheng86a}, $E_z$ should be $E_x$.}
\end{itemize}
The general solution is given by the sum of a particular solution plus an homogeneous one. \cite{cheng86a} propose a possible solution for $V$, considering the particular and homogeneous parts. The particular part reads
\begin{equation}\label{eq:potential_chr86}
  V^{\rm nc}= - \frac{ \Omega B}{c r_c} x z(z-a) = - \frac{\Omega B}{c r_c} a^2 x q(q-1)~, 
\end{equation}
where $q\equiv z/a \in [0,1]$. The corresponding parallel electric field is given by:
\begin{equation}\label{eq:epar_og}
 E_\parallel^{\rm nc}= -\frac{dV^{\rm nc}}{dx} =\frac{ \Omega B}{c r_c}a^2 q(q-1)~,
\label{Epar}
\end{equation}
whereas the perpendicular component of the non-corotational electric field is

\begin{equation}\label{eq:eperp_og}
 E_\perp^{\rm nc} = -\frac{dV^{\rm nc}}{dz} =\frac{\Omega B}{c r_c} x (2z - a)~.
\end{equation}
Therefore, it is discontinuous across the boundary, since, outside the gap, Eq.~(\ref{eq:efield_induced}) holds, and $E_\perp^{nc}=0$. This implies that there is a layer of surface charge at the lower/upper boundaries ($z=0,a$) given by
\begin{equation}
 \Sigma = \frac{\Delta E_\perp}{4\pi} = \pm\frac{\Omega B}{4\pi c} \frac{xa}{2r_c}~. 
\end{equation}
This charge layer is needed to screen the electric field outside the gap. The mechanisms leading to the replenishing and stability of such a surface charge layer are only qualitatively motivated in \cite{cheng86a} (see their eq.(3.5) and the related discussion in their \S V), and currently not corroborated by any global magnetospheric simulation.

The $a^2$ dependence in Eq.~(\ref{Epar}) arises from the second order nature of the Poisson equation, and it plays a key role in estimating the gap size and electric field, as we will see below. The homogeneous part of the solution of \cite{cheng86a} is more complicated (see their Appendix), it satisfies $E_\parallel = 0$ at the inner boundary, and its value goes to zero for $x \gtrsim a$. As a consequence, the particular solution (\ref{eq:potential_chr86}) has to be taken as a possible solution not too close and not too far from the inner boundary. However, even with the assumptions and simplifications above, the proposed solution is not unique: other dependencies in the longitudinal direction could in principle be found, because the outer boundary is not specified.

The thick OG model \cite{zhang97} employs a single-value assumption for $E_\parallel$ (hereafter we drop the subscript ${\rm nc}$), by extrapolating the thin OG formula, Eq.~(\ref{eq:epar_og}), to any gap trans-field thickness $a$:

\begin{equation}\label{eq:epar_thick_og}
 E_\parallel = \frac{\Omega B(r) a^2 q(q-1)}{c r_c}~.
\end{equation}
We express $E_\parallel$ as a function of $B_\star$, $P$ and $f$ by writing,
\begin{equation}\label{eq:epar_zc97}
 E_\parallel = f^2 \chi B_\star\left(\frac{ 2\pi R_\star }{Pc}\right)^3~,
\end{equation}
where $\chi$ is
\begin{equation}
 \chi \equiv \frac{a^2q(q-1)}{R_{lc} r_c}\frac{B(r,\theta)}{B_\star}\left(\frac{R_\star}{R_{lc}}\right)^3~.
\end{equation}
This factor is implicitly set as $\chi=1$ in the work of \cite{zhang97}, which implicitly add the following approximations:
\begin{itemize}
  \item It extrapolates to any value $f\leq 1$; however, large values of $f$ can be inconsistent with the assumption that the inner boundary of the OG is the null surface {\rm bf(see the third point at the beginning of this subsection)}.
  \item They assume that $r\sim r_c \sim R_{lc}$, which is inconsistent with the geometrical approximations listed above for building the thin gap.\footnote{The relation given by Eq.~(11) of \cite{zhang97}, $E_\parallel \approx \Omega B a^2 / c r_c\approx E_\perp f^2 R_{lc}/r_c$, is not consistent with their definition $E_\perp=Br/R_{lc}$. This inconsistency goes unnoticed due to the approximation $r\sim r_c \sim R_{lc}$.}
 \item Setting $a^2q(q-1) = f^2 R_{lc}r_c$ can be a rough estimate with a possible error of a few at least.
 \item All quantities, including $f$ and $r_c$ are supposed to be constant along the magnetic field lines.
\end{itemize}
Most of these approximations, arising from the intrinsic difficulties of the problem, are usually overlooked. Therefore, we stress that one should take Eq.~(\ref{eq:epar_zc97}) as an order-of-magnitude estimate and explore predicted spectra within a given range of feasible values.

\subsection{Comparison with numerical works}\label{sec:numerical}

\begin{table*}
\begin{center}
\caption{$E_\parallel$: analytical estimate, Eq.~(\ref{eq:epar_zc97}) with $\chi=1$, and numerical maximum values. In Takata et al. (2006) $f$ is not specified.}

\begin{tabular}[ht!]{l c c c c c}
\hline
\hline
Ref. & $f$ &  $B_\star$ & $P$ & Numerical value & Analytical value \\
& & [G] & [s] & [V/m] & [V/m] \\
\hline
Fig. 3 \cite{hirotani06} & 0.047 & $1.46\times 10^{13}$ & 0.033 & $\sim 6\times 10^7$ & $2.5\times 10^8$ \\
Fig. 4 \cite{hirotani06} & 0.060 & $1.46\times 10^{13}$ & 0.033 & $\sim 9\times 10^7$ & $4\times 10^8$ \\
Fig. 4 \cite{hirotani06} & 0.100 & $1.46\times 10^{13}$ & 0.033 & $\sim 1.5\times 10^8$ & $1.1\times10^9$ \\
Fig. 9 \cite{hirotani06} & 0.039 & $2.19\times 10^{13}$ & 0.033 & $\sim 3\times 10^7$ & $2.6\times 10^8$ \\
Fig. 6 \cite{takata06}  &  $\sim 0.3$?	& $6.2\times 10^{12}$ & 0.089 & $1-3.4 \times 10^7$ & $2.4 f^2 \times 10^9$\\
\hline
\hline
\end{tabular}
\end{center}
\label{tab:hirotani_num}
\end{table*} 

The series of works by Hirotani, Takata \& collaborators aim at overcoming some of the rough analytical approximations by solving the problem numerically. These works improve on the following aspects.

\begin{itemize}
 \item The magnetic field geometry is numerically evaluated. A dipole is still considered, but the values of $B$, $r_c$, etc. are consistently calculated at each point. The Poisson equation is solved numerically, considering all the spatial dependences.
 \item The gap is not void of particles: pairs produced are consistently taken into account and their dynamics followed (Boltzmann equation) in order to evaluate the local density.
 \item The cross section of the pair production is accurately included in the Boltzmann equation.
 \item The inclination angle is properly included in the global geometry.
 \item The simulations are in 2D (meridional sections): this allows to study the effect of the trans-field dependence of pair density and the shape of the gap.
  \item The position of the inner and outer boundaries can be consistently found by considering where the created pairs totally screen the electric field. In general, the position of these boundaries is a function of magnetic field lines (e.g., Fig. 12 of \citealt{takata06}), i.e., very different from the picture of a constant-$f$ used in analytical works.
 \item Some numerical simulations \citep{hirotani06,takata06,takata08} allow currents coming from the surface (extraction of particles or polar sparks) or from the outer boundary (due to the residual $E_\parallel$ slightly outside it). The inner boundary moves inside the null surface, with a non-trivial dependence on the geometry of the magnetic field line (i.e., the inner boundary has an irregular shape).
 \end{itemize}
A main result of these simulations is that, in presence of pair production, the electric field is partially screened by the charged pairs continuously produced and separated. In order to evaluate this effect, we can compare the numerical results presented in Figs. 3 and 4 of \cite{hirotani06}, which show the numerical values of $E_\parallel$ for the employed parameters, at different heights of the gap. \cite{hirotani06} extends the calculations down to the surface, showing that $E_\parallel$ is usually very small beneath the null surface, while it approaches a constant value in the outer magnetosphere, at least when $f\ll 1$. The quadratic dependence of $E_\parallel$ on the trans-field coordinate ($z$) is confirmed.
However, there is a quantitatively important mismatch with the estimate (\ref{eq:epar_zc97}). As shown in Table~\ref{tab:hirotani_num}, $E_\parallel$ at maximum (i.e., at the center of the gap, in the outer part) is 5-10 times smaller in the numerical simulations than in the analytical estimates.

\subsection{Trans-field gap size}\label{sec:f}

\subsubsection{Inferring $f$}

In order to take the screening into account, we introduce a screening factor $\kappa_{\rm scr}$ in the formula obtained in the vacuum thick OG approximation, Eq.~(\ref{eq:epar_zc97}):
\begin{eqnarray}\label{eq:epar_screening2}
  E_\parallel & =& \kappa_{scr} \chi f^2 B_\star\left(\frac{2\pi R_\star}{P c}\right)^3 \nonumber\\
  & = & 2.76 \times 10^5 \kappa_{\rm scr} \chi  R_6^3 f^2\frac{B_{12}}{P^3}  \frac{{\rm V}}{{\rm m}}~.
\end{eqnarray}
By equating the electrodynamics estimate, Eq.~(\ref{eq:epar_screening2}), to the expressions for $E_\parallel$ needed to sustain the gap for a given $T$, we obtain
\begin{eqnarray}
  f & = & \sqrt{\frac{E_\parallel}{B_\star \kappa_{scr} \chi}\left(\frac{cP}{2\pi R_\star}\right)^3}  ~,\nonumber\\
  & \simeq & 6.0 \sqrt{\frac{E_{\parallel,7}}{B_{12} \kappa_{scr} \chi} \left(\frac{P}{R_6}\right)^3} ~,
\end{eqnarray}
The subscript $t$ below refers to the formula of $E_\parallel$ involving a fixed $T$ (independent of $P$ and $B_\star$), Eq.~(\ref{eq:epar_t}). Instead, the indices $h$ and $s$ refer to the formulae considering the bombarded hot polar cap temperature, Eq.~(\ref{eq:epar_th}) or the RCS-reflected temperature, Eq.~(\ref{eq:epar_ts}). Thus, we have three estimates of the gap size
\begin{eqnarray}
  f_t & = & a_t P^{7/6} B_{12}^{-1/2} ~, \label{eq:f_t}\\ 
  f_h & = & a_h P^{23/18} B_{12}^{-2/3} ~, \label{eq:f_h}\\
  f_s & = & a_s P^{26/21} B_{12}^{-4/7}~, \label{eq:f_s}
\end{eqnarray}
where the pre-factors are
\begin{eqnarray}
 a_t & = & 3.6 \left(\frac{\tilde{g}_r}{R_6^3 \kappa_{scr}\chi}\right)^{1/2} \left( \frac{k_\gamma}{kT{\rm [keV]}}  \right)^{2/3} \left(\frac{R_{lc}}{r_c}\right)^{1/3}~, \label{eq:a_t}\\
 a_h & = & 7.0 \left(\frac{\tilde{g}_r}{R_6^3 \kappa_{scr}\chi}\right)^{1/2} \left( \frac{k_\gamma}{{\cal K}_h}  \right)^{2/3} \left(\frac{R_{lc}}{r_c}\right)^{1/3}~, \label{eq:a_h}\\
 a_s & = & 32 \left(\frac{\tilde{g}_r}{R_6^3 \kappa_{scr}\chi}\right)^{1/2} \left( \frac{k_\gamma}{{\cal K}_s}  \right)^{2/3} \left(\frac{R_{lc}}{r_c}\right)^{1/3}~. \label{eq:a_s}
\end{eqnarray}
Eqs.~(21) and (22) of \cite{zhang97}, with $a_h=1.6$ and $a_s=5.5$, are roughly recovered by considering $k_\gamma=0.087$ (corresponding to $E_X=3~kT$, $\Psi=\pi$ and $\mu=0$ in Eq.~(\ref{eq:egamma})), ${\cal K}_h={\cal K}_s=0.5^{1/4}=0.84$ (due to $M_e=1/2$), $r=R_{lc}$ and $R_6=\chi=\kappa_{scr}=1$. The derived formulae above 
give account of the range in which the gap size is expected for a given value of $B_{\star}$ and $P$.

\subsubsection{The $P$-$\dot{P}$ diagram and the pulsar population.}

\begin{figure*}
\centering
\includegraphics[width=.4\textwidth]{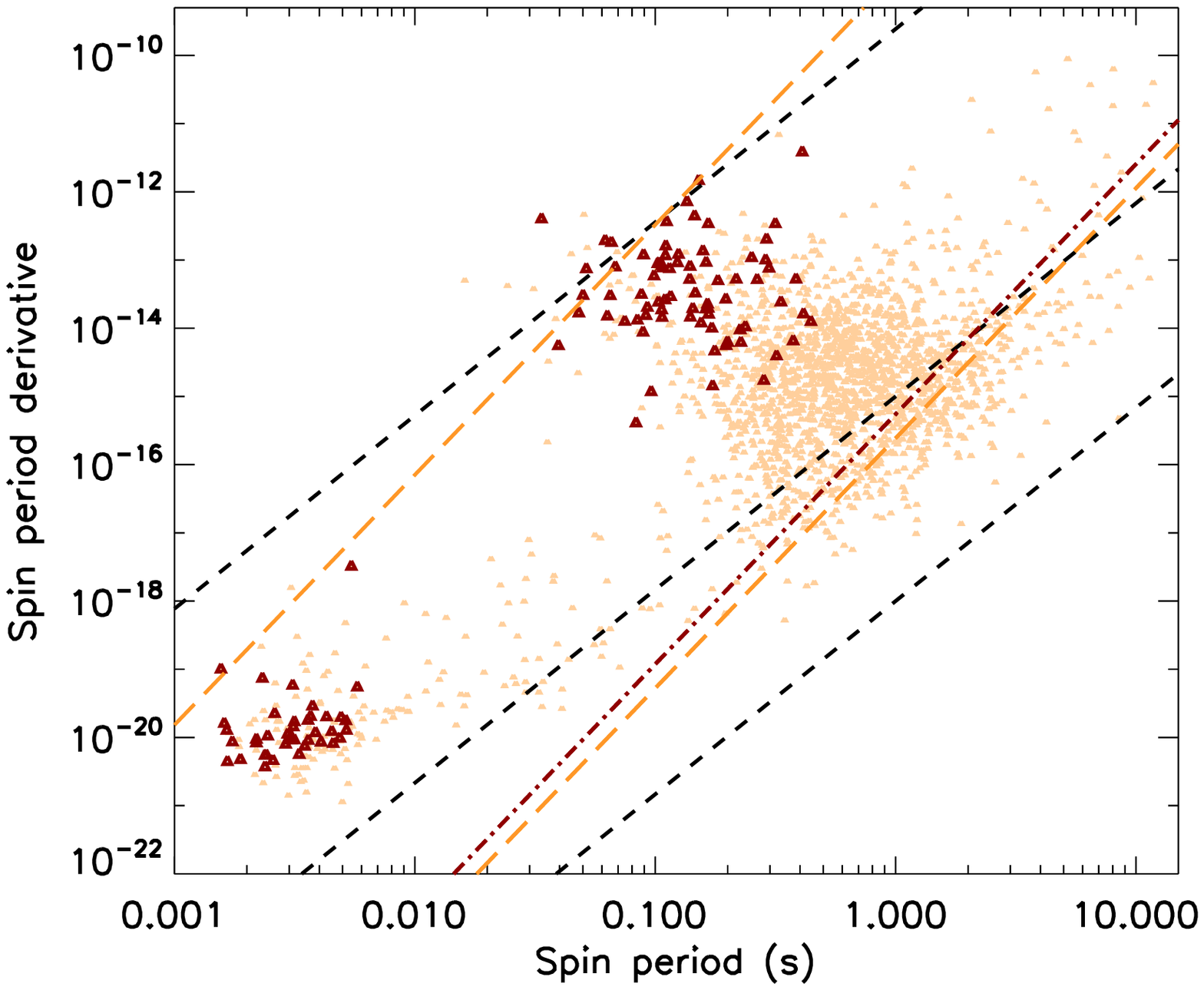}
\includegraphics[width=.4\textwidth]{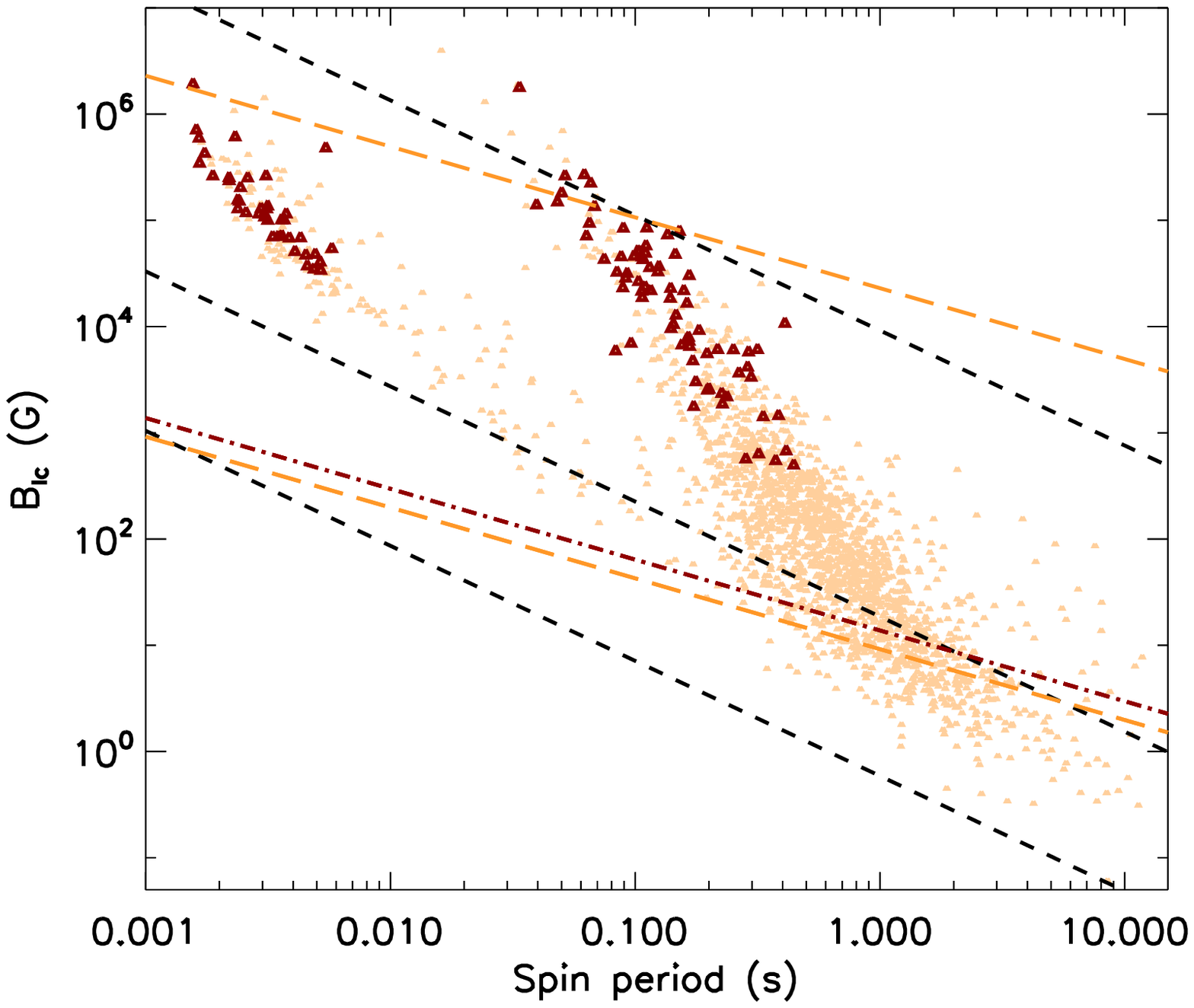}
\caption{$P$-$\dot{P}$ diagram (top) and $B_{lc}$-$P$ diagram showing all pulsars (small orange points), according to the ATNF catalog, and $\gamma$-ray pulsars (red triangles), according to the 2PC LAT catalog \citep{2fpc}. We show the lines of $f_h=1$ and $f_t=1$ for different choice (lines from top to bottom) of $a_h=0.5,1.6,100$ (black dashes) and $a_t=1,50$ (long orange dashes), covering two orders of magnitude. The consistency line of $E_{\rm hit}=E_{\rm rot}$, Eq.~(\ref{eq:hit_consistency}) is shown with a red dot-dashed line.}
 \label{fig:ppdot_f}
\end{figure*}

 Since, by definition, $f\leq 1$, the relations above would in principle estimate whether a thick OG solution exists or not for a given pulsar. Considering a $P$-$\dot{P}$ diagram, this translates into a $\gamma$-ray death-line, since, below the line $f=1$, no thick gap solution exists, i.e., the entire open magnetosphere is not able to provide enough acceleration to the particles. We revise such death-line estimates considering the various uncertainties discussed, and summarized in the pre-factors (\ref{eq:a_t})-(\ref{eq:a_s}).

In Fig.~\ref{fig:ppdot_f} we show the pulsar population according to the ATNF catalog\footnote{\url{http://www.atnf.csiro.au/research/pulsar/psrcat/}} \citep{manchester05}. The $\gamma$-ray pulsars of the second {\em Fermi}-LAT catalog \citep{2fpc} are indicated with red marks. On the left, we show their timing properties, $P$ and $\dot{P}$. From the latter, one can infer the surface dipolar magnetic field, assuming it regulates the spin-down torque of the NS, see Eq.~(\ref{eq:ppdot_spindown}):

\begin{equation}\label{eq:bdip}
 B_\star = 6.4\times 10^{19} \left(\frac{I_{45}}{h_\iota R_6^6}  P \dot{P}\right)^{1/2} ~{\rm G}~.
\end{equation}
The uncertainty of the pre-factor can be considered to be a factor $\sim 2$ (see \S 2.2 of \citealt{viganothesis} and references therein). In the right panel of Fig.~\ref{fig:ppdot_f} we also show the same population considering the magnetic field at the light cylinder, which is estimated at the equator (distance $r=R_{lc}$ to the surface) and assuming a $B\sim r^{-3}$ dependence:

\begin{equation}\label{eq:blc}
 B_{\rm lc} = B_\star\left(\frac{R_\star}{R_{lc}}\right)^3 = 5.9\times 10^8 P^{-5/2} \dot{P}^{1/2} ~{\rm G}~,
\end{equation}
where we have taken $(I_{45}/h_\iota R_6^6)^{1/2} = 1$.
In both panels we indicate lines corresponding to $f_h=1$ (black dashes), or $f_t=1$ (long cyan dashes), for three values of $a_h=0.5,1.6,100$ and $a_t=1,50$, roughly covering the expected range of variability (see Table~\ref{tab:uncertainties_og} and discussion below). The central value of $a_h=1.6$ corresponds to the value given by \cite{zhang97} and often taken as a reference.

Additionally, a dot-dashed green line is over-plotted to show a self-consistency check for the case of bombardment ($f_h=1$ lines): we require that $\dot E_{\rm hit}$, Eq.~(\ref{eq:en_hitting}) to be less than the rotational energy loss, given by\begin{eqnarray}\label{eq:erot}
 \dot{E}_{rot} = 9.6 \times 10^{30} h_\iota R_6^6 ~\frac{B_{12}^2}{P^4}~ {\rm erg~s}^{-1}~.
\end{eqnarray}
Using eqs.~(\ref{eq:en_hitting}) and (\ref{eq:erot}), the consistency condition $\dot E_{\rm hit} < \dot{E}_{rot}$ translates into:

\begin{equation}\label{eq:hit_consistency}
  B_\star > 3 \times 10^{12} ~\left[ \frac{\Delta\phi}{2\pi} \frac{f(1-f/2)M_e k_{\rm geo}^{1/3}}{ {\cal K}_\alpha R_6^3 }\right]~ P^{7/3} {\rm G}~.
\end{equation}
The violation of this condition, which happens for large periods (where the green dot-dashed line is above the black dashes) stresses that some of the assumptions in the bombardment estimates are intrinsically inconsistent (see \S\ref{sec:bombardment}).

Below the lines, i.e., for $f>1$, the thick OG predicts that no detection. While in literature it is common to find a precise death-line, corresponding to $a_h=1.6$, here we show that such line suffer a huge dispersion, for which any firm conclusion about the expected detectability of the pulsars is difficult to be drawn. The largest values of $a_h$ and $a_t$ are excluded by the presence of several pulsars below the upper lines, while the smallest values would include a large fraction of the total radio/X-ray pulsar population. Realistically speaking, even if the thick OG would apply, we should expect a progressively fading away of the sources, rather than a death line. The latter supposes an abrupt switch-off of the $\gamma$-ray mechanism: whether it is the case or not can only be studied by realistic numerical simulations.

Last, we point out that the maximum value of $f$ is, strictly speaking, constrained by the condition
that the null surface should cross the magnetic field line within the
light cylinder (see above).

\section{Discussion}\label{sec:discussion}

\begin{table*}
\begin{center}
\footnotesize
\caption{Quantities in the OG model. The impact on $E_\parallel$ is evaluated as: negligible (change by less than a factor 2), moderate (change by a factor of a few), or large (up to orders of magnitude).}
\label{tab:uncertainties_og}

\begin{minipage}{\textwidth}
\begin{tabular}[ht!]{l l l l}

\hline
\hline
Quantity	&	Assumption in \cite{zhang97}	& Estimated range	& Impact on $E_\parallel$\\
\hline
{\bf NS properties and geometry} & & & \\
Radius	& 10 km & 8-14 km & negligible (via $T_h$) \\
$B_\star(P,\dot{P})/6.4\times 10^{19} \sqrt{P\dot{P}}$ G &  1 & $\sim 0.5-2$, Eq.~(\ref{eq:bdip}) & - \\
$r_c$	 & $r_c=\sqrt{d_lR_{lc}}\sim R_{lc}$ & 0.3-2 $R_{lc}$, \S\ref{sec:rc} & large \\
$B(r) \propto r^{-b_1}$ & $b_1=3$ & $b_1 \in (2,3)$, \S\ref{sec:local_b} & moderate \\
$B(r_c) \propto r_c^{-b}$ & $b=6$ & $b \in (5,8)$, \S\ref{sec:local_b}  & moderate \\
\hline
{\bf Plasma dynamics} & & & \\
Lorentz factor $\Gamma$ in the gap & Steady-state value $\Gamma_{\rm st}$ & $\lesssim \Gamma_{\rm st}$, \cite{paper0} & moderate \\
Synchro-curvature factor $g_r$ & 1 & $\sim 1-3$, \cite{paper0} & large \\
\hline
{\bf Particle bombardment} & & & \\
Energy for bombarding particles & $10.3 P^{1/3}$ erg & 1-10 erg, \S\ref{sec:bombardment} & moderate (via $T_h$) \\
Bombardment efficiency $\epsilon_{\rm hit}$ & 1 & $\lesssim 1?$ & negligible (via $T_h$) \\
\hline
{\bf Energies} & & & \\
${\cal K}_h$ (related to $T_h$) & 0.84 & 0.5-1, Eq.~(\ref{eq:t_h_pre2}) & moderate \\
${\cal K}_s$ (related to $T_s$) & 0.84 & 0.5-1, Eq.~(\ref{eq:t_h_pre2}) & moderate \\
Cooling temperature & - & 0.05-0.2 keV & moderate \\
$k_\gamma = E_\gamma[{\rm GeV}]/kT[{\rm keV}]$ & 0.087 & 0.1-0.2 & large \\
\hline
{\bf Gap size} & & & \\
$E_\parallel(f,B,P)$ & $f^2 B_{\rm lc}$ & $\kappa_{\rm scr}\chi f^2 B_{\rm lc}$ & large \\
$\kappa_{\rm scr}$ & 1 & $\sim 0.1-0.5$ & large \\
$\chi$ & 1 & $\sim 0.1-10$ & large \\
$a_h$ & 1.6 & 0.5-100 & - \\
$a_s$ & 5.5 & 2-500 & - \\
$a_t$ & - & 0.2-70 & - \\ 
\hline
\hline
\end{tabular}

\end{minipage}
\end{center}
\end{table*}

In Table~\ref{tab:uncertainties_og}, we summarize our results, i.e., the main parameters entering in the OG models 
and their expected range of uncertainty, as well as the impact on the value of $E_\parallel$.
We stress that the only directly accessible quantities are the $\gamma$-ray spectra, the timing properties ($P$ and $\dot{P}$), and, in some cases, the X-ray flux.
Some properties of the NS star, like the radius, the temperature and the magnetic field geometry at a given age, rely on theoretical models (equation of state, MHD equilibrium of the proto-NSs, long-term cooling and magnetic field evolution). While most models of equation of state predict a radius of $\sim 8-14$ km, the range of uncertainty for the magnetic field configuration and temperature is much larger.
The geometry of the magnetic field determines the local values of $B$ and $r_c$, and the position of the inner boundary of the gap. Usually, the vacuum dipolar approximation is used, but one should keep in mind that  rotation causes an important deviation from this solution. Moreover, the inclination angle affects the magnetospheric geometry and, consequently, the position of the gap, which, in turn, affects the radius of curvature $r_c$. The presence of multipoles affect the particle dynamics close to the surface, while a large-scale twist of the magnetic field lines can also affect the outer region.

Given the uncertainties of both the exact position of the gap and the magnetospheric configuration, the exact value of $r_c$ along the gap is expected to cover a range of about one order of magnitude (see \S\ref{sec:rc}). It is a key parameter, since it regulates the radiative losses, and, in turn, the particle dynamics (see \citealt{paper0} for details on the latter).
As discussed in detail in \S\ref{sec:local_b}, it is useful to parametrize the radius of curvature and the radial dependence of $B$ within the values expected for the split monopole and the vacuum dipole solution (higher multipoles do not influence the outer magnetosphere).

If the gap is sustained by photon-photon pair production, as assumed in the thick OG model, then X-rays play a key role. They can be thermally emitted from the surface due to either the slow release of the internal residual heat, or due to the polar cap bombardment of returning currents. To estimate the second mechanism, one has to evaluate the amount of kinetic energy released at the surface. In both cases, the value of the surface temperature can be estimated within a factor of a few, if we consider the uncertainties discussed in \S\ref{sec:bombardment} and summarized in Table~\ref{tab:uncertainties_og}.

A side consideration is the evaluation of the gap trans-field thickness, $f$. It relies on an extrapolation of the estimate for $E_\parallel$ given by an analytical treatment of the thin OG, as discussed in detail in \S\ref{sec:efield_f}. We conclude that the classical estimate of the latter, which consider $\kappa_{\rm scr}=\chi=1$, overestimate by up to one order of magnitude the real value. Altogether, the uncertainties in the pre-factors, Eqs.~(\ref{eq:a_t})--(\ref{eq:a_s}), are realistically one to two orders of magnitude.

\begin{table*}
\begin{center}
\footnotesize
\caption{Ingredients in the OG model.}
\label{tab:problems_og}

\begin{minipage}{\textwidth}
\begin{tabular}[ht!]{l l l}

\hline 
\hline 
Description  & Analytical approach & Numerical approach \\
\hline
{\bf Geometry} && \\
Magnetospheric geometry		& Split monopole, dipole, twist 			& Global force-free solution \\
Boundaries of the gap 		& Null surface of dipole, light cylinder 	& Global force-free solution \\
Size of the gap 				& Free parameter 					& Global-scale pair production \\
Electrodynamics of the gaps 	& $E_\parallel$ with screening factor		& Global resistive solution \\
\hline
{\bf Plasma properties} && \\
Particle density 			& GJ value with a multiplicity/screening factor & Solution to Boltzmann equation \\
Pitch angle and $\Gamma$ distributions	& Free parameter 					& Simulations of particle dynamics \\
\hline
{\bf X-rays} & & \\
Bombardment polar cap, $T_h$ & Estimate $E_{\rm kin \star}$ & Simulations of particle dynamics \\
Surface emission 			& X-ray observed/free parameter & Cooling models \\
RCS reflection to the surface 	& Consider it only against IR/optical photons & RCS simulations \\
\hline
{\bf Pair production} & & \\
$\gamma\gamma$ pair production & Typical ranges of $E_X$ and $E_\gamma$ & Global simulation \\
$\Gamma$ and pitch angle at creation 		& Free parameter		& Boltzmann equations \\
\hline
\hline

\end{tabular}
\end{minipage}
\end{center}
\end{table*} 

In Table~\ref{tab:problems_og}, we summarize the possible analytical and numerical approaches to reduce or parametrize the uncertainties. In general, numerical studies of the gap electrodynamics overcome the single-value approximations of the photon energies. They also give realistic values for the pair production rates, considering the whole spectral distribution of both X and $\gamma$ rays. These kinds of works have been already extensively discussed (e.g., \citealt{hirotani06,takata06}), with some limitations which are not easy to overcome:
\begin{itemize}
 \item the gap is not electro-dynamically connected with the force-free magnetosphere (for instance, by imposing a continuous $E_\perp$ across the gap);
 \item the purely dipolar geometry is not a realistic prescription for the outer magnetosphere;
 \item the values of $f$ (and the upper/lower boundaries), and the outer boundary, are often fixed and treated as model parameters;
 \item the results are quite sensitive to some parameters of the models, like the inclination angle and the amount of currents flowing into the gap from outside, which are hardly constrainable a-priori;
 \item the OG is confined to the closed magnetosphere; however, it would be interesting to explore the OG model also outside the light cylinder;
\end{itemize}
A substantial improvement would be to link global magnetospheric numerical studies with the gap electrodynamics, like done in a first attempt by \cite{chen14}. A smooth match between the force-free region and the partially screened gap would help to understand the shape, size of the gap, and the electromagnetic and particle dynamics within it. Second, globally evaluated quantities, like $r_c$ and $B$, would be directly available, taking into account all the effects of rotation, and, eventually, twists and multipoles.

Some of the parameters usually taken in the literature are slightly outside the range we expect. In particular, in Paper II we will show how $k_\gamma$ should be larger than the value usually assumed, since the latter considers a vanishing cross-section of the $\gamma$-$\gamma$ interaction. Similarly, the parameters $\chi$ and $\kappa_{\rm scr}$, which strongly affects the formula for $E_\parallel$, are usually not considered in the thick OG models.

With these considerations, and considering the uncertainties discussed above, we conclude that:

\begin{itemize}
  \item In the extensive literature about the OG, many assumptions and single-value approximations are usually taken for granted or overlooked.
  \item Several of these have a large impact on the determination of the electric field accelerating particles, and from there onwards, in the predicted spectra.
  \item Special attention should be paid to the magnetospheric geometry, which affects the model in many ways (radius of curvature, local $B$, etc.).
  \item The long-term cooling temperature should be taken into account as a primary mechanism to sustain the gap, rather than appealing to an unlikely fully efficient conversion into soft X-ray photons by RCS scattering.
  \item The latter should in any case consider the  Doppler shift, and assuming a more realistic non-total reflection to the NS surface. 
  \item The extrapolation from the thin OG model to the thick OG relies on a vast number of approximations and assumptions, some of which are not mutually consistent;
  \item in particular, the widely used formulae for $E_\parallel(f,B,P)$, Eq.~(\ref{eq:epar_zc97}), should be taken as a very rough estimate: it relies on many approximations and single-values. We note that it overestimates the numerically-obtained values by a factor $\sim 5-10$ (see Table~\ref{tab:hirotani_num}).
\end{itemize}

The ranges of uncertainties for all the ingredients of the OG model have been summarized here. One of the most rough approximations is perhaps that of considering a mono-energetic choice of the $\gamma$-ray energy, $E_\gamma$. In order to improve this aspect, we need to study the pair production process which regulates the interplay between plasma and photons. Another important issue is the variation of the parameters along the gap length. Both issues will be analytically studied in Paper II, where we also will address the impact of the uncertainties on the predicted $\gamma$-ray spectra and how such modified predictions relate to observational data.

\section*{Acknowledgements}

This research was supported by the grant AYA2012-39303 and SGR2014-1073 (DV, DFT) and the Project Formosa TW2010005, for bilateral research between Taiwan and Spain (KH, DFT). KH is partly supported by the Formosa Program between National Science Council in Taiwan and Consejo Superior de Investigaciones Cientificas in Spain administered through grant number NSC100-2923-M-007-001-MY3. 
The research leading to these results has also received funding from the
European Research Council under the European Union's Seventh Framework
Programme (FP/2007-2013) under ERC grant agreement 306614 (MEP). MEP
also acknowledges support from the Young Investigator Programme of the Villum Foundation.
We are grateful to the referee for a careful reading and the useful comments.

\appendix

\section{Estimates of radius of curvature and local magnetic field values}\label{sec:magnetosphere}

In the OG models, the vacuum dipole is usually taken as the background geometry, from which one can derive exact formulae for the radius of curvature, and for the location of the inner boundary (i.e., the null surface), assuming some inclination angle. However, the magnetosphere is expected to deviate considerably from such naive picture, especially in the outer regions, where the rotation cause the lines to become twisted and stretched. In this Appendix, we recall some introductory picture of the standard force-free magnetosphere, and consider how we can estimate the relations between the radius of curvature, the distance along a magnetic field line, and the local values of the magnetic field.

\subsection{Force-free magnetosphere}

The standard picture of the magnetosphere is largely based on the model developed  by \citep{goldreich69}. With the hypothesis of a free supply of plasma able to replenish the magnetosphere, the configuration is supposed to be force-free, because any net force would be quickly compensated by a reorganization of the highly conducting plasma in order to give zero net force. If the plasma is a perfect conductor and rigidly co-rotates with the star, with rotational velocity $\vec{v}_{rot}=\vec{\Omega}\times\vec{r}$, the unipolar induction generates an electric field
\begin{equation}\label{eq:efield_induced_app}
\vec{E}=-\frac{\vec{v}_{rot}}{c}\times \vec{B} = - \frac{\vec{\Omega}\times\vec{r}}{c}\times \vec{B}~.
\end{equation}
The {\em Goldreich-Julian (GJ) charge density} is obtained from Maxwell's equation as:

\begin{equation}\label{eq:rho_gj_app}
  \rho_{gj}(r,\theta) = \frac{\vec{\nabla}\cdot\vec{E}}{4\pi} = -\frac{\vec{\Omega}\cdot\vec{B}}{2\pi c(1-(\Omega r \sin\theta/c)^2)}~,
\end{equation}
where hereafter we work in spherical coordinates $(r,\theta,\phi$). The rotation induces the separation of charges. For a magnetic dipole with magnetic moment parallel to the angular velocity vector, the polar regions are negatively charged, while the equatorial region is positively charged. The {\it null surfaces} are defined by $\vec{\Omega}\cdot\vec{B}=0$.

Plasma can rigidly co-rotate with the star only in a region spatially limited by the finiteness of the speed of light. The {\em light cylinder} is the distance from the rotational axis at which a co-rotating particle would reach the speed of light:
\begin{equation}\label{eq:def_light_cyl}
  R_{lc}=\frac{c}{\Omega}=4.77\times 10^9~P[\mbox{s}]~{\rm cm} = 10^8\frac{P}{0.021 {\rm s}}~{\rm cm}~,
\end{equation}
which, for the range of pulsars periods, $P\sim 10^{-3}-10$~s, corresponds to tens to tens of thousands of stellar radii ($R_\star \sim 10^6$ cm). Some magnetic field lines close inside the light cylinder, while those connected to the polar region cross it. The {\em separatrix} is the line dividing the {\em co-rotating magnetosphere} (closed field lines) from the open field lines region. The polar cap is defined as the portion of the surface connected with the open field lines.

Despite the simplicity of the problem, i.e., the description of a rotating magnetized sphere surrounded by perfectly conducting plasma, the solutions are not trivial. As a matter of fact, the rotation induces azimuthal currents, $J_\phi$, which in turn modify the magnetic field, especially close to the light cylinder. Close to the surface, $r\ll R_{lc}$, if the magnetosphere is not twisted ($B_\phi=0$), then $J_\phi$ is negligible, and the magnetic field is potential, i.e., $\vec{\nabla}\times \vec{B}=0$. However, in the outer magnetosphere, where $r\lesssim R_{lc}$, the rotational current $J_\phi$ becomes important and the deviations from any vacuum solution cannot be neglected. As a consequence, the electro-dynamical description of the outer regions, and the open field lines in particular, is not trivial. There, the magnetic field is thought to be mainly radial and twisted.

\subsection{The ideal vacuum dipole and the polar cap}\label{sec:vacuum}

A potential magnetic field consists, in general, by a superposition of different multipoles with degree $l$, each one with a radial dependence $\sim r^{-(l+2)}$. Far from the surface, the dipole ($l=1$) is the dominant component and it is the most usual assumed configuration:
\begin{eqnarray}\label{eq:dipole_field}
&& \Gamma=B_\star R_\star^2\frac{\sin^2\theta}{2r}~,\\
&& B_r=B_\star\cos\theta\left(\frac{R_\star}{r}\right)^3 ~,\\
&& B_\theta=B_\star\frac{\sin\theta}{2}\left(\frac{R_\star}{r}\right)^3 ~,
\end{eqnarray}
where we have introduced the surface magnetic field (at the pole), $B_\star$, and the magnetic flux function $\Gamma$, which is related to the poloidal component of the magnetic field by

\begin{equation}
 \vec{B}_{pol} = \frac{\vec{\nabla}\Gamma \times \hat{\phi}}{r\sin\theta}~.
\end{equation}
For a given magnetic configuration, the functional form of $\Gamma(r,\theta)$ effectively represents the definition of a magnetic field line, since, along it, $\Gamma$ is constant by definition. For a vacuum dipole, then, a magnetic field line is defined by $r \propto \sin^2\theta$, where the proportionality constant depends on which line we are considering. In particular, the {\em separatrix} (i.e., the line marking the boundary between open and closed lines) can be identified by imposing that, at the equator $\sin\theta=1$, $r=R_{lc}$, so that:
\begin{equation}\label{eq:separatrix}
 \sin\theta = \sqrt{\frac{r}{R_{lc}}}~. 
\end{equation}
The surface ($r=R_\star$) footprint of the separatrix gives us the semi-opening angle of the polar cap:
\begin{equation}\label{eq:pc_angle}
  \sin\theta_{pc}=\sqrt{\frac{R_\star}{R_{lc}}}=\sqrt{\frac{R_\star\Omega}{c}}~,
\end{equation}
which, for $R_\star = 10$ km, corresponds to
\begin{equation}\label{eq:pc_angle2}
  \theta_{\rm pc} \sim \frac{0.8^\circ}{\sqrt{P[\mbox{s}]}}~,
\end{equation}
which means a polar cap radius
\begin{equation}\label{eq:pc_radius}
 r_{\rm pc} \sim R_\star\sqrt{\frac{2\pi R_\star}{Pc}}  \sim  \frac{145}{\sqrt{P[\mbox{s}]}} {\rm m}~.
\end{equation}
The GJ charge density for an aligned rotator, Eq.~(\ref{eq:rho_gj_app}), is
\begin{equation}\label{eq:rho_gj_close}
  \rho_{gj}(r,\theta)\simeq-\frac{\Omega B_\star (1-(3/2)\sin\theta^2)}{2\pi c}~.
\end{equation}
In this case, the null surfaces are defined by $|\sin\theta_\pm|=\sqrt{2/3}$, which means $\theta_\pm=55^\circ,125^\circ$. 
Combined with Eq.~(\ref{eq:separatrix}), this gives us the null surface along the separatrix: 
\begin{equation}\label{eq:null_sep}
[r,\sin\theta_{\rm null}](\iota=0)=\left[\frac{2}{3}R_{lc},\sqrt{\frac{2}{3}}\right]~.
\end{equation}
In a general, non-axisymmetric case, i.e., for an oblique dipole rotator with inclination angle $\iota\neq 0$, the angle of the null surface is given by \citep{cheng00}
\begin{eqnarray}\label{eq:null_general}
  && \sin^2\theta_{\rm null}(\iota,\phi) = \nonumber\\
  && = \frac{1}{2} + \frac{1/3 \pm \tan\iota\cos\phi[(\tan\iota\cos\phi)^2 + 8/9]^{1/2}}{2[1+(\tan\iota\cos\phi)^2]}~.
\end{eqnarray}
For an orthogonal rotator, $\iota=90^\circ$, then $\sin\theta=1/\sqrt{2}$. For intermediate values, there is a non trivial dependence of the null surface on $\phi$.

\subsection{The split monopole}

The split monopole solution \citep{michel73} is an analytical force-free configuration with a smooth matching across the light cylinder. The components of the magnetic field are:
\begin{eqnarray}\label{split_monopole_components}
&& B_r=\pm B_0\left(\frac{R_\star}{r}\right)^2 ~,\\
&& B_\theta=0 ~,\\
&& B_\varphi= B_0\frac{R_\star^2}{r R_{lc}}\sin\theta~.
\end{eqnarray}
The magnetic field is directed outwards in one hemisphere and inwards in the other one, in order to preserve $\vec{\nabla}\cdot\vec{B}=0$. In the equatorial plane, the discontinuity of $B_r$ in the $\theta$-direction implies a toroidal current sheet $J_\varphi$. The magnetic field lines are twisted, and the angle between the toroidal and radial magnetic field components is $\arctan(r\sin\theta/R_{lc})$. The current is purely radial and proportional to the angular velocity $\Omega$:

\begin{equation}
 J_r=\frac{B_0 R_\star^2 \Omega}{2\pi}\frac{\cos\theta}{r^2}~,
\end{equation}
which means that particles move only radially close to the speed of light. To understand this result, consider the co-rotating frame, in which particles move only along the twisted field lines. Seen from an inertial observer, the toroidal component of this velocity is exactly compensated by the azimuthal drift of the magnetic field lines due to rigid rotation, thus $J_\phi=0$. Despite its simplicity, this solution is thought to be the appropriate geometry for the regions close to the light cylinder and beyond it. The main feature is the slow decrease with radius (compared with the $r^{-3}$ dependence for a dipole), and the stretching and twisting of the lines.

\subsection{Rotating dipolar solutions}

\cite{contopoulos99} numerically found a force-free solution for an aligned rotating dipole, smooth across the light cylinder. This solution has been later confirmed by several other works \citep{goodwin04,contopoulos05,mckinney06,spitkovsky06,chen14}. Compared with a non-rotating dipole, the magnetic field lines are inflated by the rotation, and a larger fraction (+36$\%$ in \citealt{contopoulos99}) of the magnetic flux is contained in the open field line region. As a consequence, a larger fraction of magnetic field lines goes through the light cylinder, the polar cap is larger, and the separatrix is more stretched than in the vacuum dipole. In that region, and in the region beyond the light cylinder, the lines are twisted, and the pattern is remarkably similar to the split monopole discussed above.

Another characteristic feature of such numerical solutions is the so-called Y-point that defines the boundary between open and closed field lines. At the Y-point, the separatrix displays a kink, and it lies on the equatorial plane, close to the light cylinder. Beyond the Y-point, an equatorial current sheet separates the northern and southern hemispheres, where radial and azimuthal fields point to opposite directions, like in the split-monopole configuration. \cite{spitkovsky06} first numerically solved the time-dependent oblique rotator in 3D. It maintains the main features of the aligned rotator solution, with the main difference that the equatorial current sheet oscillates around the rotational equatorial plane. The particle-in-cell simulations by \cite{chen14} show that these current sheets could host the $\gamma$-ray production.

In conclusion, the effects of rotation are to bring the magnetosphere to a configuration similar to a vacuum dipole, close to the surface, and to a split monopole beyond the light cylinder. Any fine-tuned attempt to prescribe a geometry is dependent on some assumptions: the presence of higher multipoles and/or currents (i.e., line twists) circulating in the magnetosphere (deviation from vacuum solution), the relative inclination between $\vec{B}$ and $\vec{\Omega}$. As a consequence, we take the vacuum dipole and the split monopole as two extremes between which we can evaluate the expected range of some OG model parameters, like the radius of curvature and the local value of $B$.


\subsection{Distance along a magnetic field line}\label{sec:distance_line}

\begin{figure}
\centering
\includegraphics[width=.45\textwidth]{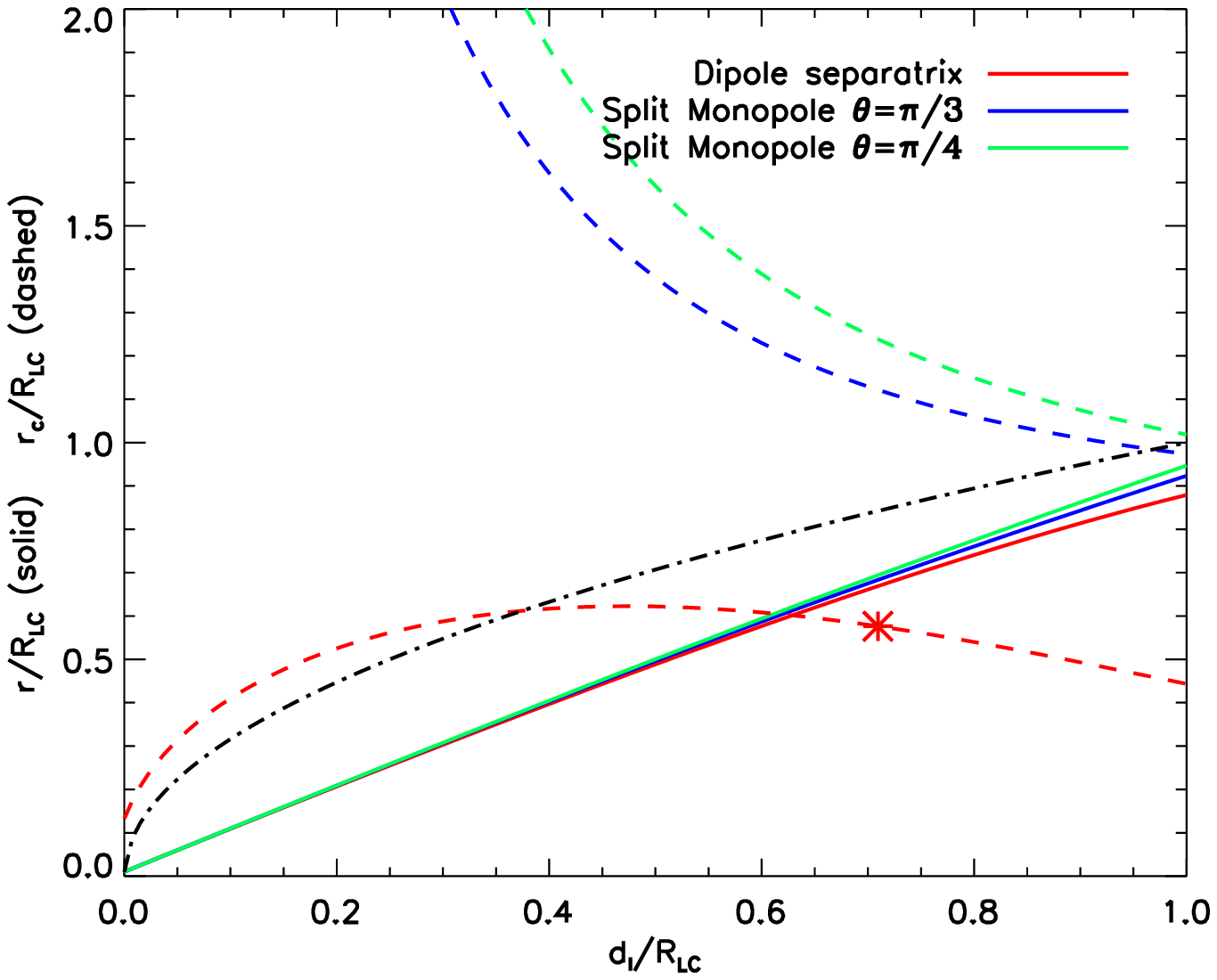}
\caption{Functions $r(d_l)$ (solid lines) and $r_b(d_l)$ (dashed), for the vacuum dipole separatrix (red), and for two magnetic field lines (blue, $\theta=\pi/3$, and green, $\theta=\pi/4$) of the split monopole. The dot-dashed black line represent the approximation $r_c\sim (d_lR_{lc})^{1/2}$. All quantities are in units of $R_{lc}$.}
 \label{fig:distance_line}
\end{figure}

In order to find the distance $d_l$ between two points along a magnetic field line, we have to integrate the infinitesimal displacement along the line, which, in spherical coordinates, is defined by:
\begin{equation}
 \delta l = \sqrt{\delta r^2 + r^2 \delta \theta^2 + r^2\sin^2\theta \delta\phi^2}~.
\end{equation}
Here $\delta r$, $r\delta \theta$, $r\sin\theta\delta \phi$ represent the infinitesimal displacements along the line in the three directions, which are involved in the definition of a magnetic field line:
\begin{equation}
 \frac{\delta r}{B_r}=\frac{r\delta\theta}{B_\theta}=\frac{r\sin\theta\delta\phi}{B_\phi}~,
\end{equation}
so that, between two points $l_1$ and $l_2$ of the same field line, we have
\begin{equation}
 d_l=\int_{l_1}^{l_2} \left[1 + \left(\frac{B_\theta}{B_r}\right)^2 + \left(\frac{B_\phi}{B_r}\right)^2 \right]^{1/2}   {\rm d} r~.
\end{equation}
On the other hand, for a specified functional form of the potential $\Gamma(r,\theta)$, we can impose a relation between $r$ and $\theta$, and give a prescription for $l_1(r,\theta)$ and $l_2(r,\theta)$.

For an aligned vacuum dipole, 
eqs.~(\ref{eq:dipole_field}), 
the distance along the magnetic field line between the surface and a point at a radial distance $r$ is 
\begin{equation}\label{eq:xr_dipole}
 d_l(r) = \int_{R_\star}^{r} \left[1 + \left(\frac{\sin\theta(r)}{2\cos\theta(r)}\right)^2\right]^{1/2}   {\rm d} r~.
\end{equation}
In order to calculate the distance along the line, we have to specify which line we are considering, i.e. the relation $\theta(r)$.
For lines closer to the pole, with $\sin\theta\rightarrow 0$, then $d_l \sim r$, because lines are almost purely radial.  For highly curved lines, and with $r\gg R_{lc}$, $d_l$ is larger than $r$. For the separatrix, along which the OG is located, Eq.~(\ref{eq:separatrix}) holds, thus
\begin{equation}
  d_{\rm sep}(r) = \int_{R_\star}^{r} \left[\frac{4R_{lc} - 3r}{4(R_{lc}-r)}\right]^{1/2}   {\rm d} r~.
\end{equation}
If $r\ll R_{lc}$, then $d_{\rm sep} \sim r$. Note that the equation above is strictly valid for the separatrix in the vacuum dipole approximation. For an oblique rotator, the separatrix surface is not axisymmetric, thus non-trivial dependences on the azimuthal angle $\phi$ will appear in Eq.~(\ref{eq:xr_dipole}). For a split monopole, lines are mostly radial and twisted, and, being $B_\theta=0$, a line is simply defined by a constant $\theta$. The function $d_l(r)$ scales with $R_{lc}$ (with corrections of the order $R_\star/R_{lc}\ll 1$).

In Fig.~\ref{fig:distance_line} we show with solid lines the function $r(d_l)$ for the separatrix of the dipole, and for different lines of the split monopole. The approximation $d_l \sim r$ holds, except in the outer magnetosphere in the limit case of vacuum dipole, where $d_l \lesssim 1.3 r$. The value of $d_l$ is used whenever an integration along a magnetic field line is needed, e.g., in solving the equation of motion in \S\ref{sec:bombardment}.

\subsection{Radius of curvature}\label{sec:rc}

%

For a given magnetic field, the curvature vector of the magnetic field lines is defined as:
\begin{equation}
 \vec{\kappa}_b= - \hat{b}\times (\vec{\nabla}\times\hat{b}) = (\hat{b}\cdot\vec{\nabla})\hat{b}~,
\end{equation}
where $\hat{b}=\vec{B}/B$, and we have used the vectorial identities.

Since particles drift along the rotating field lines, their velocity vector, seen by a distant observer, does not coincide with the magnetic field vector, instead:
\begin{eqnarray}\label{eq:v_curv}
 && v_r=c f_v\frac{B_r}{B}~, \nonumber\\
 && v_\theta=c f_v\frac{B_\theta}{B}~, \nonumber\\
 && v_\phi=c \left(f_v\frac{B_\phi}{B} + \frac{r\sin\theta}{R_{lc}}\right)~, \nonumber\\
 && f_v^\pm= - \frac{B_\phi}{B}\frac{r\sin\theta}{R_{lc}} \pm \sqrt{1-\left(\frac{r\sin\theta}{R_{lc}}\right)^2\frac{B_r^2+B_\theta^2}{B^2}}~.
\end{eqnarray}
The upper sign of $f_v$ corresponds to the outward motion along the magnetic field line, whereas the lower sign to the inward motion \citep{hirotani11}. The curvature vector characterizing the motion of the particles is then

\begin{equation}
 \vec{\kappa}= - \hat{v}\times (\vec{\nabla}\times\hat{v}) = (\hat{v}\cdot\vec{\nabla})\hat{v}~.
\end{equation}
The inverse of $|\vec{\kappa}|$ is the radius of curvature of the particle trajectories:

\begin{equation}\label{eq:def_rc}
 r_c =||(\hat{v}\cdot\vec{\nabla})\hat{v}||^{-1}~.
\end{equation}
It differs from the radius of curvature of the magnetic field lines, $r_b=||\vec{\kappa}||^{-1}$, if rotation is important, i.e., close to the light cylinder. Otherwise, a first approximation are the analytical formulae for $r_b$. For a vacuum dipole, it is:

\begin{equation}\label{eq:rc_dip}
 r_b^{\rm dip} = \frac{4}{3}r\frac{[1-(3/4)\sin^2\theta]^{3/2}}{\sin\theta[1-(1/2)\sin^2\theta]}~.
\end{equation} 
This function grows linearly with $r$ and monotonically decreases with $\sin\theta$, being larger at the poles (where lines are more stretched). In the limit of $\sin\theta\rightarrow 0$ (axis), $r_b^{\rm dip}\rightarrow 4r/3\theta$, whereas, in the limit $\theta\rightarrow \pi/2$ (equator), $r_b^{\rm dip}\rightarrow r/3$.

For the separatrix, considering Eq.~(\ref{eq:separatrix}), we can express $r_c^{\rm dip}$ as
\begin{equation}\label{eq:rcsep}
  r_b^{\rm sep} \sim \frac{4}{3}\sqrt{rR_{lc}} \frac{[1-3r/4R_{lc}]^{3/2}}{1-r/2R_{lc}}~.
\end{equation}
In Fig.~\ref{fig:distance_line} we show $r_c^{\rm sep}$ with a  as a function of distance along the line, $d_l$. Note that \cite{zhang97} use the approximation $r_c = \sqrt{d_lR_{lc}}$ (dot-dashed black line), which is a good approximation (apart from a factor $4/3$) for the low-latitude gaps (the PC gaps, see eqs.~5-6 and related discussion in \citealt{arons79}). The asterisk indicates the position of the null surface (for an aligned rotator), usually assumed as the inner boundary. In the outer regions, such approximation fails.

In a split monopole, the calculation of $r_b$ provides

\begin{equation}\label{eq:rc_mon}
 r_b^{\rm mon} = \frac{R_{lc}}{\varpi} \frac{(1 + \varpi^2)^{3/2}}{\left[(2 + \varpi^2)^2 + (\varpi^4 + \varpi^2)\frac{\cos^2\theta}{\sin^2\theta} \right]^{1/2}}~,
\end{equation}
where $\varpi\equiv r\sin\theta/R_{lc}$. For $\varpi\gg 1$ (wind zone, where the solution makes sense), $r_c^{\rm mon}\sim k_m(\theta) r$, with $k_m(\theta)$ function of $\theta$. Compared with the dipolar case, the important qualitative difference lies in the radial dependence, which is not monotonic. Actually, $r_b^{\rm mon}$ decreases with $r$ up to a $\theta$-dependent value of $r > R_{lc}$, because of the increasing twist of the lines outward ($B_\phi$ decreases slower than $B_r$). In other words, the outflowing currents cause the lines to be more and more curved.



Even considering the inward or outward particles ($f_v^+$ or $f_v^-$ in Eq.~\ref{eq:v_curv}), the analytical formulae (\ref{eq:rc_dip}) and Eq.~(\ref{eq:rc_mon}) hold close to the surface. In the outer region, however, the rotation contributes to substantially modify the radius of curvature: $f_v^\pm$ deviates from unity, and $r_c$ can change by a factor of several.

In conclusion, we expect the radius of curvature to have a value between the dipole and monopole approximations, with dependences on the gap position and size, the global geometry of the magnetic field and the inclination angle. Since we do not know the exact magnetospheric configuration, we treat the radius of curvature as a free parameter, with values between $r_c\sim 0.3-2~R_{lc}$.

The radius of curvature is the more important parameter to be considered in the radiative losses of any gap model. In paper II, we will use it as the fundamental integration variable that characterizes the dynamics along the gap.

\subsection{Local magnetic field}\label{sec:local_b}

An important parameter in the OG model is the local value of $B$. It is usually related to the surface value, $B_\star$, by assuming the dependence $B(r) = B_\star(R_\star/r)^3$, which roughly holds for a dipole. In a split monopole, such dependence would be softer, with $B(r) \sim r^{-2}$. Note that these simple power-law dependences are not approximate, since they do not consider the angular dependencies. Also, a single power-law index is not expected to accurately describe the dependence across the whole magnetosphere.

Therefore, since the local values are highly geometry-dependent, we will parametrize the unknown magnetic field geometry by means of an effective power index $b_1$, used to define the local $B$ by

\begin{equation}
 B(r)=B_\star \left(\frac{R_\star}{r}\right)^{b_1}~.
\end{equation}
Since the outer magnetosphere is supposed to be qualitatively a transition between a dipole and a split monopole, then we will consider $b_1 \sim 2-3$.

If one is interested in the local $B$ as a function of $r_c$, and we assume that
\begin{equation}
 r=R_{lc} \left(\frac{r_c}{R_{lc}}\right)^{b_2}~,
\end{equation}
then
\begin{equation}
 B(r_c)=B_\star \left(\frac{R_\star}{R_{lc}}\right)^{b_1} \left(\frac{r_c}{R_{lc}}\right)^{-b}~,
\end{equation}
where $b=b_1b_2$. Close to the polar cap, $b_2\sim 2$ and $b_1 \sim 3$, then $b\sim 6$. In the outer magnetosphere, however, $b_2$ is larger and $b_1$ is slightly smaller. Precise values can be obtained from numerical force-free configurations, thus in Paper II we choose to explore $b \sim 5-8$.

\bibliography{og}

\end{document}